\documentclass{aa} 
\usepackage{graphicx}
\usepackage{natbib,twoopt}
\usepackage[breaklinks=true]{hyperref} 
\bibpunct{(}{)}{;}{a}{}{,} 
\makeatletter
\newcommandtwoopt{\citeads}[3][][]{\href{http://adsabs.harvard.edu/abs/#3}%
{\def\hyper@linkstart##1##2{}%
\let\hyper@linkend\@empty\citealp[#1][#2]{#3}}}
\newcommandtwoopt{\citepads}[3][][]{\href{http://adsabs.harvard.edu/abs/#3}%
{\def\hyper@linkstart##1##2{}%
\let\hyper@linkend\@empty\citep[#1][#2]{#3}}}
\newcommandtwoopt{\citetads}[3][][]{\href{http://adsabs.harvard.edu/abs/#3}%
{\def\hyper@linkstart##1##2{}%
\let\hyper@linkend\@empty\citet[#1][#2]{#3}}}
\newcommandtwoopt{\citeyearads}[3][][]%
{\href{http://adsabs.harvard.edu/abs/#3}
{\def\hyper@linkstart##1##2{}%
\let\hyper@linkend\@empty\citeyear[#1][#2]{#3}}}
\makeatother

\usepackage{txfonts}
\usepackage{color}
\begin{document}
%
\title{Impact of photoevaporative mass loss on masses and radii of water-rich sub/super-Earths}

   \author{K. Kurosaki,
          \inst{1}
          M. Ikoma\inst{1}
          \and
          Y. Hori\inst{2}
          }

   \institute{Department of Earth and Planetary Science, The University of Tokyo, 7-3-1 Hongo, Bunkyo-ku, Tokyo 113-0033, Japan; 
              \email{kkurosaki@eps.s.u-tokyo.ac.jp}
         \and
             Division of Theoretical Astronomy, National Astronomical Observatory of Japan, 2-21-1
             Osawa, Mitaka, Tokyo 181-8588, Japan
                          }

   \date{Submitted 11 July 2013 / Accepted 3 December 2013}

 
  \abstract
   {Recent progress in transit photometry opened a new window to the interior of super-Earths.
From measured radii and masses,  {we can infer constraints on planetary internal compositions.}
It has been recently revealed that super-Earths orbiting close to host stars (i.e., hot super-Earths) are diverse in composition. This diversity is thought to arise from diversity in volatile content.
}
   {The stability of the volatile components, which we call the envelopes, is to be examined, because hot super-Earths, which are exposed to strong irradiation, undergo photo-evaporative mass loss. While several studies investigated the impact of photo-evaporative mass loss on hydrogen-helium envelopes, there are few studies as to the impact on water-vapor envelopes, which we investigate in this study. To obtain theoretical prediction to future observations, we also investigate the relationships among masses, radii, and semi-major axes of water-rich super-Earths and also sub-Earths that have undergone photo-evaporative mass loss.}
   {We simulate the interior structure and evolution of highly-irradiated sub/super-Earths that consist of a rocky core surrounded by a water envelope which include mass loss due to the stellar XUV-driven energy-limited hydrodynamic escape.}
   {We find that the photo-evaporative mass loss has a significant impact on the evolution of hot sub/super-Earths. With a widely-used empirical formula for XUV flux from typical G-stars and the heating efficiency of 0.1 for example, the planets of less than 3 Earth masses orbiting 0.03~AU have their water envelopes completely stripped off.
   We then derive the threshold planetary mass and radius below which the planet loses its water envelope completely as a function of the initial water content and find that there are minimums of the threshold mass and radius.}
   {We constrain the domain in the parameter space of planetary mass, radius, and the semi-major axis in which sub/super-Earths never retain water envelopes in 1-10~Gyr. This would provide an essential piece of information for understanding the origin of close-in, low-mass planets. The current uncertainties in stellar XUV flux and its heating efficiency, however, prevent us from deriving robust conclusions. Nevertheless, it seems to be a robust conclusion that Kepler planet candidates contain a significant number of rocky sub/super-Earths.
}

   \titlerunning{Impact of mass loss for water-rich sub/super-Earths}
\authorrunning{Kurosaki, Ikoma \& Hori~2013}
\maketitle

%
%

\section{Introduction}
Exoplanet transit photometry opened a new window to the interior and atmosphere of exoplanets. 
The biggest advantage of this technique would be that planetary radii are measured, while planetary masses are measured via other techniques, such as the radial velocity method and the transit timing variation method. 
Measured mass and radius relationships help us infer the internal structure and bulk composition of exoplanets theoretically, which give crucial constraints to formation and evolution processes of the planets. 
A growing number of small-sized exoplanets with radii of 1 to 2~$R_\oplus$ have been identified,
which are often referred to as super-Earths \citep{Batalha2013}. 
Also, planet candidates detected by the Kepler space telescope include sub-Earth-sized objects, such as Kepler-20~e \citep{Fressin2011}, Kepler-42~b, c, d \citep{Muirhead2012}, and Kepler-37~b, c \citep{Barclay2013}. 
We can thus discuss the compositions of such small planets to gas giants by comparing theory with current observations. 

Transiting super-Earths detected so far show a large variation in radius, suggesting diversity in composition. 
There are many theoretical studies on mass-radius relationships for planets with various compositions and masses  \citep{Valencia2007, Fortney2007, Sotin2007, Seager2007, Grasset2009, Wagner2011, Swift2012}.
A recent important finding, which compares theory to observation is that there are a significant number of low-density super-Earths that are larger in size than they would be if they were rocky. 
This implies that  {these transiting super-Earths possess components less dense than rock. 
From a viewpoint of planet formation, the possible components are hydrogen-rich gas and water, which make an outer envelope. 
A small fraction of H-rich gas or water is known to be enough to account for observed radii of the low-density super-Earths \citep{Adams2008, Valencia2010}. 

The stability of the envelopes are, however, to be examined. 
Transiting planets are generally orbit close to their host stars (typically $\lesssim 0.1$~AU), because detection probability of planetary transits is inversely proportional to the semi-major axis (e.g., \citealt{Kane2007}).
These close-in planets are highly irradiated and exposed to intense X-ray and  {ultraviolet radiation} (hereafter XUV) that come from their host stars. 
This causes the planetary envelope to escape hydrodynamically from the planet (e.g., \citealt{Watson1981}). 
This process is often called the photoevaporation of planetary envelopes. 
As for massive close-in planets, namely, hot Jupiters, the possibility of the photoevaporation and its outcome have been investigated well both theoretically and observationally (e.g., \citealt{Yelle2008} and references therein).  

While the photoevaporation may not significantly affect the evolution and final composition of hot Jupiters except for extremely irradiated or inflated hot Jupiters, its impact on small close-in planets in the sub/super-Earth mass range should be large, partly because their envelope masses are much smaller than those of hot Jupiters. 
For example, \citet{Valencia2010} investigated the structure and composition of the first transiting super-Earth CoRoT-7~b and discussed the sustainability of the possible H+He envelope with a mass of less than 0.01~\% of the total planetary mass. 
The envelope mass was consistent with its measured mass and radius. 
The estimated lifetime of the H+He envelope was, however, only 1 million years, which was much shorter than the host star's age (2-3~Gyr). 
This suggests that CoRoT-7~b is unlikely to retain the H+He envelope at present. 

Young main-sequence stars are known to be much more active and emit stronger XUV than the current Sun (e.g., \citealt{Ribas2005}). Therefore, even if a super-Earth had a primordial atmosphere initially, it may lose the atmosphere completely during its history.  
These discussions concerning the photo-evaporative loss of H+He envelopes were done for GJ~1214~b \citep{Nettelmann2011, Valencia2013}, super-Earths orbiting Kepler-11 \citep{Lopez2012, Ikoma2012}, and CoRoT-7~b \citep{Valencia2010}. 
Systematic studies were also done by \citet{Rogers2011}
and \citet{Lopez2013}.
Those studies demonstrated the large impact of the photoevaporation on the stability of H+He envelopes for super-Earths.  
In particular, \citet{Lopez2013} performed simulations of coupled thermal contraction and photo-evaporative mass loss of rocky super-Earths with H+He envelopes. They found that there were threshold values of planetary masses and radii, below which H+He envelopes were completely stripped off.
\citet{Owen2013} also performed similar simulations with detailed consideration of the mass loss efficiency for an H+He envelope based on \citet{Owen2012}. They argued that evaporation explained the correlation between the semi-major axes and planetary radii (or planet densities) of KOIs.

In this study, we focus on water-rich sub/super-Earths.
Planet formation theories predict that low-mass planets  {migrate toward their host star, which is strongly supported by the presence of many close-in super-Earths, from cooler regions (e.g., \citealt{Ward1986})
where they may have accreted a significant amount of water.}
This suggests that water/ice-rich sub/super-Earths may also exist close to host stars. 
Therefore, similar discussions should be done for water envelopes of close-in super-Earths. 
However, there are just a few studies, which treat specific sub/super-Earths such as CoRoT-7~b \citep{Valencia2010} and Kepler-11~b \citep{Lopez2012}.
No systematic study is yet to be done for the stability of water envelopes.  

The purpose of this study is, thus, to examine the stability of primordial water envelopes of close-in sub/super-Earths against photo-evaporation. 
To this end, we simulate the thermal evolution of planets with significant fractions of water envelopes (i.e., water-worlds), incorporating the effect of stellar-XUV-driven photo-evaporative mass loss. 
The theoretical model is described in section~2. 
 {As for the atmosphere model, the details are described in Appendix A.} 
In section~3, we show the evolutionary behavior of the water-rich planets. 
Then, we find threshold values of planetary masses and radii below which such water-rich planets are incapable of retaining primordial water envelopes for a period similar to ages of known exoplanet-host stars (i.e., 1--10~Gyr). 
In section~4, we compare the theoretical mass-radius distribution of water-rich planets 
with that of known transiting planets.
Furthermore,
we compare the threshold radius with sizes of Kepler objects of interest (KOIs) to suggest that KOIs include a significant number of rocky planets.
Finally, we summarize this study in section~5. 

%
%

\section{Numerical models} \label{model}
In this study, we simulate the evolution of the mass and radius of a planet that consists of water and rock, including the effects of mass loss due to photoevaporation. 
The structure model is depicted in Fig.~\ref{Fig_interior}.
The planet is assumed to consist of three layers in spherical symmetry and hydrostatic equilibrium: namely, from top to bottom,
it consisted of a water vapor atmosphere, a water envelope, and a rocky core. At each interface, the pressure and temperature are continuous. 

The assumptions and equations that determine the planet's interior structure and thermal evolution are described in section~\ref{equations} and section~\ref{evolution}, respectively. 
The equations of state for the materials in the three layers are summarized in section~\ref{eos}. 
The structure of the atmosphere and the photoevaporative mass loss, both of which govern the planet's overall evolution, are described in section~\ref{atmosphere}  {(see also Appendix~A)} and section~\ref{loss}, respectively.
Since a goal of this study is to compare our theoretical prediction with results from transit observations, we also calculate the transit radius, which is defined in section~\ref{transit}. Finally, we summarize our numerical procedure in section~\ref{procedure}. 

\subsection{Interior structure}\label{equations}
The interior structure of the planet is determined by the differential equations (e.g. \citealt{Kippenhahn1990}),
\begin{eqnarray}
\frac{\partial P}{\partial M_r} &=& -\frac{GM_r}{4\pi r^4}, \label{IE01} \\
\frac{\partial r}{\partial M_r} &=& \frac{1}{4\pi r^2 \rho}, \label{IE02}\\
\frac{\partial T}{\partial M_r} &=& -\frac{GM_r T}{4\pi r^4 P}\nabla, \label{IE03}
\end{eqnarray}
and the equation of state,
\begin{eqnarray}
\rho &=& \rho (P, T), \label{IE04}
\end{eqnarray}
where $r$ is the planetocentric distance, 
$M_r$ is the mass contained in the sphere with radius of $r$,
$P$ is the pressure, $\rho$ is the density, $T$ is the temperature, and 
$G$ (= $6.67 \times 10^{-8}$~dyne~cm$^2$~g$^{-2}$) is the gravitational constant.
The symbol $\nabla$ is the temperature gradient with respect to pressure. 
We assume that the water envelope and rocky core are fully convective and the convection is vigorous enough that the entropy $S$ is constant; namely, 
\begin{equation}
   \nabla = \left( \frac{\partial\ln T}{\partial\ln P} \right)_{S}. \label{IE+02}
\end{equation}

Equations~(\ref{IE01}), (\ref{IE02}), and (\ref{IE03}) require three boundary conditions. 
The inner one is $r = 0$ at $M_r = 0$. 
The outer boundary corresponds to the interface between the envelope and the atmosphere, which is called the tropopause. 
The tropopause pressure $P_{\rm{ad}}$ and temperature $T_{\rm{ad}}$ are determined from the atmospheric model; the details of which is described in section~\ref{atmosphere} and Appendix A. 
The atmospheric mass is negligible, relative to the planet total mass $M_p$. 
In our calculation, the atmospheric mass is less than 0.1~\% of the planetary mass.
Thus, the outer boundary conditions are given as
\begin{equation}
 \begin{array}{lllll}
	P = P_{\rm{ad}} & \mbox{and} & T = T_{\rm{ad}} & \mbox{at} & M_r = M_p. \label{Bd01}
 \end{array}
\end{equation}
As mentioned above, the pressure and temperature are also continuous at the interface between the water envelope and the rocky core. 
\begin{figure}[htbp]
	\begin{center}
	\resizebox{\hsize}{!}{\includegraphics{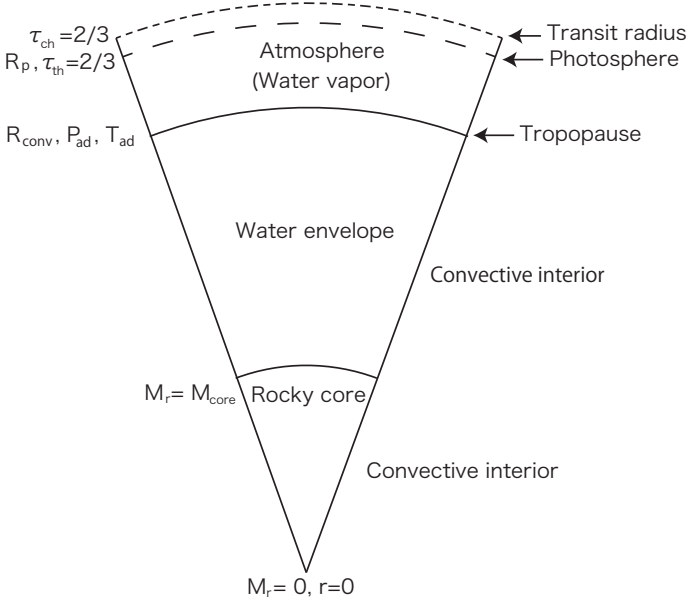}}
	\caption{Model of the planetary structure in this study.}
	\label{Fig_interior}
	\end{center}
\end{figure}

\subsection{Thermal evolution}\label{evolution}
The thermal evolution of the planet without internal energy generation is described by (e.g., \citealt{Kippenhahn1990})
\begin{equation}
   \frac{\partial L}{\partial M_r} = -T \frac{\partial S}{\partial t}, \label{Ec04}
\end{equation}
where $L$ is the intrinsic energy flux passing through the spherical surface with radius of $r$, $S$ is the specific entropy, and $t$ is time. 
Since the entropy is constant in each layer, the integrated form of Eq.~(\ref{Ec04}) is written as
\begin{equation}
-L_{p} = \frac{\partial \bar{S}_e}{\partial t} \int_{M_c}^{M_{p}} T dM_r 
             +\frac{\partial \bar{S}_c}{\partial t} \int_{0}^{M_{c}} T dM_r , \label{Ec05}
\end{equation}
where
$L_{p}$ is the total intrinsic luminosity of the planet, $M_c$ is the mass of the rocky core, and $\bar{S}_e$ and $\bar{S}_c$ are the specific entropies in the water envelope and the rocky core, respectively. 
In integrating Eq.~(\ref{Ec04}), we have assumed $L=0$ at $M_r=0$.

In the numerical calculations of this study, we use 
the intrinsic temperature $T_{\rm{int}}$, instead of $L_p$,
which is defined by
\begin{equation}
T_{\rm{int}}^{4} \equiv  \frac{L_{p}}{4\pi R_{p}^{2}\sigma}, \label{Ec06}
\end{equation}
where $R_p$ is the planet photospheric radius (see section~\ref{atmosphere} for the definition) and $\sigma$ is the Stefan-Boltzmann constant
(= $5.67 \times 10^{-5}~\rm{erg}~\rm{cm}^{-2}~\rm{K}^{-4}~\rm{s}^{-1}$).

\subsection{Equation of state (EOS)}\label{eos}
In the vapor atmosphere, the temperature and pressure are sufficiently high and low, respectively, so that the ideal gas approximation is valid. 
We thus adopt the ideal equation of state, incorporating the effects of dissociation of H$_2$O. 
In practice, we use the numerical code developed by \citet{Hori2011}, which calculate chemical equilibrium compositions among H$_2$O, H$_2$, O$_2$, H, O, H$^+$, O$^+$ and e$^-$.

At high pressures in the water envelope, the ideal gas approximation is no longer valid, because pressure due to molecular interaction is not negligible.
In this study, we mainly use the water EOS H$_2$O-REOS \citep{Nettelmann2008}, which contains the ab initio water EOS data at high pressures of \citet{French2009}.
H$_2$O-REOS covers a density range from $1.0\times 10^{-6}~\rm{g~cm}^{-3}$ to $15~\rm{g~cm}^{-3}$ and a temperature range from $1.0\times10^3$~K to $2.4\times 10^{4}$~K. 
For $T$ and $\rho$ outside the ranges that H$_2$O-REOS covers, we use SESAME 7150 \citep{SESAME}.

The rocky core is assumed to be mineralogically the same in composition as the silicate Earth. 
We adopt a widely-used EOS and the Vinet EOS, and calculate thermodynamic quantities following \citet{Valencia2007}. 

\subsection{Atmospheric model}\label{atmosphere}
As described above, we consider an irradiated, radiative-equilibrium atmosphere on top of the water envelope. 
The thermal properties of the atmosphere govern the internal structure and evolution of the planet. 
To integrate the atmospheric structure, we follow the prescription developed by \citet{Guillot2010} except for the treatment of the opacity. 
Namely, we consider a semi-grey, plane-parallel atmosphere in local thermal equilibrium. 
The wavelength domains of the incoming (stellar) and outgoing (planetary) radiations are assumed to be completely separated; the former is visible, while the latter is near or mid infrared.

We solve the equation of radiative transfer by integrating the two sets (for incoming and outgoing radiations) of the zeroth and first-order moment equations for radiation with the Eddington's closure relation: 
the incoming and outgoing radiations are linked through the equation of radiative equilibrium (see Eqs.~[10]--[11] and [17]--[19] of \citet{Guillot2010}). 
\citet{Guillot2010} derived an analytical, approximate solution, which reproduced the atmospheric structure from detailed numerical simulations  {of hot Jupiters} (see also \citet{Hansen2008}) well.
The solution depends on opacities in the visible and thermal domains.   
\citet{Guillot2010} also presented empirical formulae for the mean opacities of solar-composition
(i.e., hydrogen-dominated) gas. 

However, no empirical formula is available for opacities of water vapor of interest in this study.
We take into account the dependence of the water-vapor opacity on temperature and pressure and integrate the momentum equations numerically. 
The details about the mean opacities and momentum equations are described in Appendix A. 

The bottom of the atmosphere is assumed to be the interface between the radiative and convective zones. 
We use the Schwarzschild criterion (e.g., see \citealt{Kippenhahn1990}) to determine the interface. 
The pressure and temperature at the interface  ($P_{\mathrm{ad}}$, $T_{\mathrm{ad}}$) are used as the outer boundary conditions for the structure of the convective water envelope.

The photospheric radius $R_p$ used in Eq.~(\ref{Ec06}) is the radius at which the thermal optical depth measured from infinity, $\tau$, is 2/3; namely,
\begin{equation}
	\tau=\int_{R_p}^{\infty}\kappa_{\mathrm{th}}^{\mathrm{r}}\rho dr = \frac{2}{3},
	\label{AT05}
\end{equation}
where $\kappa_\mathrm{th}^{r}$ is the Rosseland mean opacity for the outgoing radiation (see Appendix~A for the definition).  
This level is above the tropopause, the radius of which is written by $R_{\mathrm{conv}}$ (see Fig.\ref{Fig_interior}).
We evaluate the atmospheric thickness $z$ (= $R_{p}-R_{\mathrm{conv}})$ by integrating the hydrostatic equation
from $P = P_{\mathrm{ad}}$ to $P=P_{\rm{ph}}$ using
\begin{equation}
z =- \int_{P_{\mathrm{ad}}}^{P_{\rm{ph}}} \frac{dP}{g\rho} 
= -\int_{P_{\mathrm{ad}}}^{P_{\rm{ph}}} \frac{\mathcal R}{\mu g} \frac{T}{P}dP, \label{AT03}
\end{equation}
where $g$ is the constant gravity, $\mathcal R$ (= 8.31$\times 10^{7} $ erg K$^{-1}$ g$^{-1}$) is the gas constant, and $\mu$ is the mean molecular weight.
$P_{\rm{ph}}$ is the photospheric pressure that we calculate by integrating
\begin{equation}
\frac{dP}{d\tau} = \frac{g}{\kappa_{\mathrm{th}}^{\rm{r}}} \label{AT06a}
\end{equation}
from $\tau=0$ to $2/3$.

\subsection{Mass loss}\label{loss}
The mass loss is assumed to occur in an energy-limited fashion. 
Its rate, including the effect of the Roche lobe, is given by \citep{Erkaev2007}
\begin{equation}
\dot{M} =-\frac{\varepsilon F_{\mathrm{XUV}}R_{p}\pi R_{\mathrm{XUV}}^2 }{GM_p K_{\mathrm{tide}}}, \label{ML01}
\end{equation}
where
$\varepsilon$ is the heating efficiency,
which is defined as the ratio of the rate of heating that results in hydrodynamic escape to that of stellar energy absorption;
$F_{\mathrm{XUV}}$ is the incident flux of X-ray and UV  {radiation} from the host star, 
$K_{\mathrm{tide}}$ is the potential energy reduction factor due to stellar tide;
and $R_{\rm{XUV}}$ is the effective radius at which the planet receives the incident XUV flux.
In Eq.~(\ref{ML01}), we have assumed $R_{\rm{XUV}}=R_{p}$,
which is a good approximation for close-in planets of interest \citep{Lammer2013}.
%
It is noted that \citet{Lammer2013} focused on the hydrogen-helium atmosphere. Since
the scale height of the vapor atmosphere is smaller than that of a hydrogen-helium atmosphere with the same temperature,
$R_{\rm{XUV}}\simeq R_{p}$ is a good approximation also for the vapor atmosphere.

In this study, we suppose that the host star is a G-star and adopt the empirical formula derived by \citet{Ribas2005} for $F_{\mathrm{XUV}}$:
\begin{eqnarray}
F_{\rm{XUV}} = \left\{
\begin{array}{lc}
504 \left( \displaystyle \frac{a}{1\mathrm{AU}} \right)^{-2} \mathrm{erg}~\mathrm{s}^{-1}~\mathrm{cm}^{-2} &(t<0.1\rm{Gyr}) \\ \label{ML04}
29.7\left(\displaystyle \frac{t}{1\mathrm{Gyr}} \right)^{-1.23} \left( \displaystyle \frac{a}{1\mathrm{AU}} \right)^{-2} \mathrm{erg}~\mathrm{s}^{-1}~\mathrm{cm}^{-2} & (t\ge0.1\rm{Gyr}). \label{ML02}
\end{array}
\right.
\end{eqnarray}

We use the formula for $K_{\mathrm{tide}}$ derived by \citet{Erkaev2007},
\begin{equation}
K_{\mathrm{tide}} = \frac{(\eta-1)^2(2\eta+1)}{2\eta^3}, \label{ML05}
\end{equation}
where $\eta$ is the ratio of the Roche-lobe (or Hill) radius
to the planetary radius, $R_p$.

The value of the heating efficiency is uncertain, because minor gases such as CO$_2$ contribute to it via radiative cooling.
For photoevaporation of hot-Jupiters, $\varepsilon$ is estimated to be on the order of 0.1 (\citet{Yelle2008} and reference therein).
Thus, we adopt $\varepsilon=0.1$ as a fiducial value and investigate the sensitivity of our results to $\varepsilon$.

Finally, we assume that the rocky core never evaporates.
That is simply because
we are interested in the stability of water envelopes in this study.
Whether rocky cores evaporate or not is beyond the scope of this study.

\subsection{Transit radius}\label{transit}
The planetary radius measured via transit photometry is different from the photospheric radius defined in the preceding subsection.
The former is the radius of the disk that blocks the stellar light ray that grazes the planetary atmosphere in the line of sight.
This radius is called the transit radius hereafter in this paper.
Below we derive the transit radius, basically following \citet{Guillot2010}.
Note that \citet{Guillot2010} assumed the plane-parallel atmosphere,
while we consider a spherically symmetric structure,
because the atmospheric thickness is not negligibly small relative to the planetary radius in some cases in this study.

We first introduce an optical depth that is called the chord optical depth, $\tau_{\mathrm{ch}}$
(e.g. \citealt{Guillot2010}). 
The chord optical depth is defined as
\begin{equation}
\tau_{\mathrm{ch}}(r,\nu) = \int_{-\infty}^{+\infty} \rho\kappa_{\nu} ds, \label{TR01}
\end{equation}
where $r$ is the planetocentric distance of the ray of interest (see Fig.\ref{TRfig_01}),
$s$ is the distance along the line of sight measured from the point where the line is tangent to the sphere,
and $\kappa_{\nu}$ is the monochromatic opacity at the frequency $\nu$.
Using $\tau_{\rm{ch}}$, we define the transit radius, $R_{\mathrm{tr}}$, as
\begin{equation}
\tau_{\mathrm{ch}}(R_{\mathrm{tr}}) = \frac{2}{3}. \label{TR+00}
\end{equation}

Let the altitude from the sphere of radius $r$ be $z_{\mathrm{tr}}$.
Then
$s^2 = (r+z_{\rm{tr}})^2 - r^2$ (Fig.\ref{TRfig_01}).
Eq.(\ref{TR01}) is written as
\begin{equation}
\tau_{\mathrm{ch}}(r,\nu) = 2\int_{0}^{\infty} \rho\kappa_{\nu}\frac{z_{\mathrm{tr}}+r}{\sqrt{z_{\mathrm{tr}}^2+2rz_{\mathrm{tr}}}}dz_{\mathrm{tr}}. \label{TR02}
\end{equation}
Furthermore
for convenience, we choose pressure $P$ as the independent variable, instead of $z_{\mathrm{tr}}$.
Using the equation of hydrostatic equilibrium,
\begin{equation}
\frac{dP}{dz_{\mathrm{tr}}} = - \frac{GM_p\rho}{(r+z_{\mathrm{tr}})^2}, \label{TR-01}
\end{equation}
one obtains
\begin{equation}
\tau_{\mathrm{ch}}(\nu,r) =-\frac{2}{g_r} \int_{P_r}^{0} \kappa_{\nu} \frac{(1+z_{\rm{tr}}/r)^3}{\sqrt{(1+z_{\rm{tr}}/r)^{2}-1}} dP, \label{TR03}
\end{equation}
where
\begin{equation}
g_r = \frac{GM}{r^2} \label{TR-04b}
\end{equation}
and $P_r$ is the pressure at $r$.
To integrate Eq.(\ref{TR03}), we write
$z_{\mathrm{tr}}$
as a function of $P$.
To do so, we integrate Eq.(\ref{TR-01}) and obtain
\begin{equation}
\int_{0}^{z_{\mathrm{tr}}}\frac{dz'}{(r+z')^2} = - \int_{P_r}^{P_z} \frac{dP}{GM_p\rho}, \label{TR-02}
\end{equation}
where $P_z$ is the pressure at $z_{\mathrm{tr}}$.
Eq.(\ref{TR-02}) is integrated as
\begin{eqnarray}
\frac{1}{r+z_{\mathrm{tr}}} &=& \frac{1}{r} - \frac{1}{r^2g_r} \int_{P_z}^{P_r} \frac{dP}{\rho} \nonumber \\
&=& \frac{1}{r} - \frac{z_p(P_r, P_z)}{r^2},
\label{TR-03}
\end{eqnarray}
where
\begin{equation}
z_p(P_r, P_z) \equiv \int_{P_z}^{P_r} \frac{P}{\rho g_r}d\ln P. \label{TR-04a}
\end{equation}
Thus, $z$ is written as
\begin{equation}
z_{\mathrm{tr}} = z_p \left( 1-\frac{z_p}{r} \right)^{-1}. \label{TR-05}
\end{equation}

Note that $z_p$ corresponds to the altitude in the case of a plane-parallel atmosphere
and $(1-z_p/r)^{-1}$ is the correction for spherical symmetry.

\begin{figure}[htbp]
\begin{center}
\resizebox{\hsize}{!}{\includegraphics{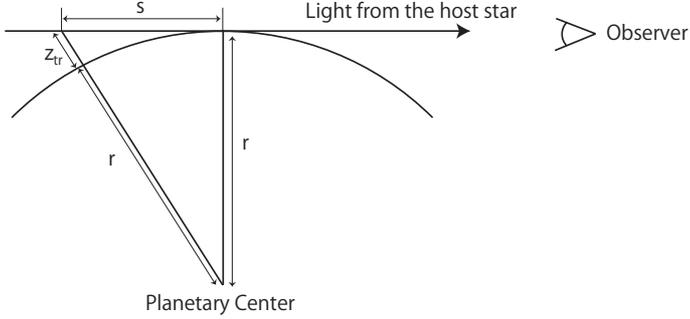}}
\caption{The concept of the chord optical depth.}
\label{TRfig_01}
\end{center}
\end{figure}

\subsection{Numerical procedure}\label{procedure}

To simulate the mass and radius evolution simultaneously, we integrate Eqs.~(\ref{Ec05}) and (\ref{ML01}) by the following procedure. 

First, we simulate two adiabatic interior models that are separated in time by a time interval $\Delta t$ for the known $M_p (t)$ and an assumed $M_p (t+ \Delta t)$. 
To be exact, the two structures are integrated for two different values of $T_\mathrm{int}$. In doing so, we integrate Eqs.~(\ref{IE01})-(\ref{IE04}) inward from the tropopause to the planetary center, using the fourth-order Runge-Kutta method. 
The inward integration is started with the outer boundary condition given by Eq.~(\ref{Bd01}); $P_\mathrm{ad}$ and $T_\mathrm{ad}$ are calculated according to the atmospheric model described in section~\ref{atmosphere}. 
We then look for the solution that fulfills the inner boundary condition (i.e., $r = 0$ at $M_r = 0$) in an iterative fashion. 
Note that determining $P_\mathrm{ad}$ and $T_\mathrm{ad}$ requires the gravity in the atmosphere (or $R_\mathrm{conv}$), which is obtained after the interior structure is determined. Thus, we have to find the solution in which the interior and atmospheric structures are consistent with each other also in an iterative fashion. 

Then we calculate $\Delta t$ from the second-order difference equation for Eq.~(\ref{Ec05}), which is written as 
\scriptsize
\begin{equation}
	\Delta t = - \frac{
		[ \bar{S}_e (t+\Delta t) - \bar{S}_e (t)  ]
		[ \Theta_e (t+\Delta t) + \Theta_e (t) ]
		+
		[ \bar{S}_c (t+\Delta t) - \bar{S}_c (t)  ]
		[ \Theta_c (t+\Delta t) + \Theta_c (t) ]
		}
	{L_p (t+\Delta t) + L_p (t)},
	\label{NM01}
\end{equation}
\normalsize where 
\begin{equation}
	\Theta_e (t) \equiv \int^{M_p (t)}_{M_c} T (t) dM_r, 
	\hspace{3ex}
	\Theta_c (t) \equiv \int^{M_c}_{0} T (t) dM_r.
	\label{NM02}
\end{equation}
Using this $\Delta t$, we integrate Eq.~(\ref{ML01}) to calculate $M_p  (t + \Delta t)$ as 
\begin{equation}
	M_p (t + \Delta t) = M_p (t) + \dot{M} \Delta t.
	\label{NM03}
\end{equation}
The assumed $M_p (t + \Delta t)$ is not always equal to that obtained here. 
Therefore the entire procedure must be repeated until the $M_p (t+\Delta t)$ in Eq.~(\ref{NM03}) coincides with that assumed for calculating Eq.~(\ref{NM01}) with satisfactory accuracy, which is $\lesssim 0.1$~\% in our simulations. 

Once we obtain the interior and atmospheric structure, we calculate the transit radius by the procedure described in section~\ref{transit}. Finally, we have confirmed that our numerical code reproduces the mass and radius relationship for super-Earths well which is presented by \citet{Valencia2010}. 

%
%

\section{Mass evolution}\label{MLresult}
In this section, we show our numerical results of the mass evolution of a close-in water-rich planet.
The evolution is controlled by the following five parameters: the initial total mass of the planet ($M_{p,0}$), the initial luminosity ($L_{0}$), the initial water mass fraction ($X_{\rm{wt},0}$), the semi-major axis ($a$), and the heating efficiency ($\varepsilon$). Below, we adopt $L_{0}=1\times 10^{24}$~erg~s$^{-1}$, $X_{\rm{wt},0}=75$~\%, $a=0.1$~AU, and $\varepsilon=0.1$ as fiducial values unless otherwise noted. 
We also show how the five parameters affect the fate of a close-in water-rich planet.

\subsection{Examples of mass evolution} \label{MLRESLT1}
Figure \ref{MT_water} shows examples of the mass evolution for water-rich planets with six different initial masses {; $L_{0}=1\times 10^{24}$~erg~s$^{-1}$, $X_{\rm{wt},0}=75$~\%, $a=0.1$~AU, and $\varepsilon=0.1$ in these simulations, as stated above}. The smallest planet loses its water envelope completely in 1~Gyr (the dashed line), while more massive planets retain their water envelopes for 10~Gyr (solid lines). This means that a water-rich planet below a threshold mass ends up as a naked rocky planet. 

The presence of such a threshold mass is understood in the following way.
Using Eq.(\ref{ML01}), we define a characteristic timescale of the mass loss ($\tau_{M}$) as
\begin{equation}
\tau_{M} = \left| \frac{X_{\rm{wt}}M_p}{\dot{M}_p} \right| = \frac{4GK_{\rm{tide}}X_{\rm{wt}}M_p\rho_{\rm{pl}}}{3\varepsilon F_{\rm{XUV}}}, \label{result_tm_01}
\end{equation}
where $\rho_{\rm{pl}}$ is the mean density of the planet.
As the planetary mass decreases, the mass-loss timescale becomes shorter.
This trend is enhanced by the $M-\rho$ relationship that the mean density decreases as $M_p$ decreases, according to our numerical results for water-rich planets.

In addition, the time-dependence of the stellar XUV flux (see Eq.~[\ref{ML02}]) is a crucial factor to cause a striking difference in behavior between the low-mass and high-mass planets.
Using Eq.~(\ref{ML04}), we obtain the following relation for $\tau_M$:
\begin{eqnarray}
\tau_{M} \simeq \left\{
\begin{array}{lcc}
3\times 10^8 f~~\rm{yr},
&\mathrm{for}&\mathit{t} < 0.1~\mathrm{Gyr}, 
\label{result_tm_02} \\
3\times 10^8 \displaystyle \left( \frac{t}{0.1~\rm{Gyr}}\right)^{1.23} f~~\rm{yr},
&\mathrm{for}&\mathit{t} \ge 0.1~\mathrm{Gyr}, \label{result_tm_03}
\end{array}
\right.
\end{eqnarray}
where
\begin{equation}
f = 1 
\left(\frac{a}{0.1\rm{AU}} \right)^{2} \left(\frac{X_{\rm{wt}}M_p}{M_{\oplus}} \right) 
\left(\frac{\rho_{\rm{pl}}}{0.1\rm{g}~\rm{cm}^{-3}} \right)
\left( \frac{K_{\rm{tide}}}{0.9} \right) \left(\frac{\varepsilon}{0.1}\right)^{-1}.  \label{result_tm_03a}
\end{equation}
Note that $0.1$~g~cm$^{-3}$ is a typical value of $\rho_{\rm{pl}}$ in the case of sub-Earth-mass planets with the age of $10^8$ years, according to our calculations.
As seen in Eq.(\ref{result_tm_03}), $\tau_{M}$ becomes longer rapidly with time. This implies that 
small planets that satisfy $\tau_{M}<0.1$~Gyr experience a significant mass loss. In other words, massive planets that avoid significant mass loss in the early phase hardly lose their mass for 10~Gyr.
Thus, there exists a threshold mass below which a planet never retains its water envelope for a long period. Our numerical calculations found that the threshold mass (hereafter $M_{\rm{thrs}}$) is $0.56~M_{\oplus}$ for the fiducial parameter set, which is in good agreement with $M_p<0.4~M_{\oplus}$ as derived from Eq.(\ref{result_tm_02}).

A similar threshold mass was found by \citet{Lopez2013} for H+He atmospheres of rocky planets. 
Hydrogen-rich planets are more vulnerable to the photo-evaporative mass loss than water-rich planets.
According to their study,
the threshold mass of the hydrogen-rich planet at 0.1~AU is $\sim 5~M_{\oplus}$.
That is, $M_{\rm{thrs}}$ for water-rich planets is smaller by a factor of $\sim 10$ than that of hydrogen-rich planets.
\begin{figure}[htbp]
\begin{center}
\resizebox{\hsize}{!}{\includegraphics{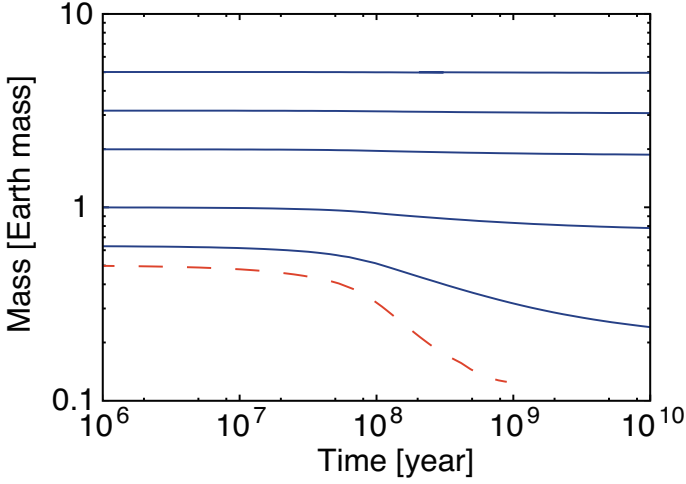}}
\caption{Mass evolution of close-in water-rich planets. The blue solid lines represent planets that retain their water envelopes for 10 Gyr. In contrast, the planet shown by the red dashed line loses its water envelope completely in 10 Gyr.
We set $L_{p,0}=1\times 10^{24}$~erg~s$^{-1}$, $X_{\rm{wt},0}=75$\%, $a=0.1$AU, and $\varepsilon=0.1$ for all the planets. In this model, we assume that the rocky core never evaporates.}
\label{MT_water}
\end{center}
\end{figure}

\subsection{Dependence on the initial planet's luminosity} \label{MLRESLT2}
The evolution during the first 0.1~Gyr determines the fate of a water-rich planet, as shown above.
Such a trend is also shown by \citet{Lopez2013} for H+He atmospheres of rocky planets.
This suggests that the sensitivity of the planet's fate to the initial conditions must be checked. In particular, the initial intrinsic luminosity may affect the early evolution of the planet significantly, because the planetary radius, which has a great impact on the mass loss rate, is sensitive to the intrinsic luminosity; qualitatively, a large $L_{0}$ enhances mass loss because of a large planetary radius. 
On the other hand, $L_0$ is uncertain, because it depends on how the planet forms
(e.g. accretion processes of planetesimals, migration processes and giant impacts).
However, as shown below, the fate of the planet is insensitive to choice of $L_{0}$

Fig.~\ref{L_Mthrs} shows $M_{\rm{thrs}}$ as a function of $L_0$ for $a=0.02, 0.03, 0.05$ and 0.1~AU.
We have found that $M_{\rm{thrs}}$ is almost independent of $L_0$.
This is because an initially-luminous planet cools down rapidly, so that the integrated amount of water loss during the high-luminosity phase is negligible.
This is confirmed by the following argument.
The mass loss, $\Delta M$, at the early stage can be estimated by
\begin{equation}
\Delta M \sim \dot{M} \tau_{\rm{KH}}, \label{result_lm_01}
\end{equation}
where $\tau_\mathrm{KH}$ is the typical timescale of Kelvin-Helmholtz contraction,
\begin{equation}
\tau_{\rm{KH}} \simeq \frac{GM_p^2}{2R_pL_p}. \label{evolv_01}
\end{equation}
With Eqs. (\ref{result_tm_01}) and (\ref{evolv_01}) given, Eq.(\ref{result_lm_01}) can be written as
\begin{eqnarray}
\Delta M &\sim& M_p \frac{\tau_\mathrm{KH}}{\tau_M}
= M_p\frac{\varepsilon}{2K_{\rm{tide}}} \cdot \frac{\pi R_p^2 F_{\rm{XUV}}}{L_p} \label{result_lm_02} \\
&\sim& 3\times 10^{-2} \left( \frac{F_{\rm{XUV}}}{504~\rm{erg}~\rm{cm}^{-2}~\rm{s}^{-1}} \right) 
\left(\frac{\varepsilon}{0.1} \right) \left( \frac{K_\mathrm{tide}}{0.9}\right)^{-1} 
\nonumber \\
&&\times \left( \frac{a}{0.1~\rm{AU}} \right)^{-2} \left( \frac{R_p}{3~R_{\oplus}} \right)^2 \left( \frac{L_p}{10^{24}~\rm{erg}~\rm{s}^{-1}} \right)^{-1} M_p. \label{result_lm_03}
\end{eqnarray}
Because $F_{\rm{XUV}}$ is constant in the early phase,
$\Delta M$ decreases as $L_p$ increases; that is, the Kelvin-Helmholtz contraction proceeds more rapidly.
Therefore, the choice of the value of $L_0$ has little effect on the total amount of water loss, as far as $L_0$ is larger than $10^{24}~\rm{erg~s}^{-1}$.
For smaller $L_0$, $R_p$ is insensitive to $L_0$.
Thus, $M_{\mathrm{thrs}}$ is insensitive to $L_0$.
\begin{figure}[htbp]
\begin{center}
\resizebox{\hsize}{!}{\includegraphics{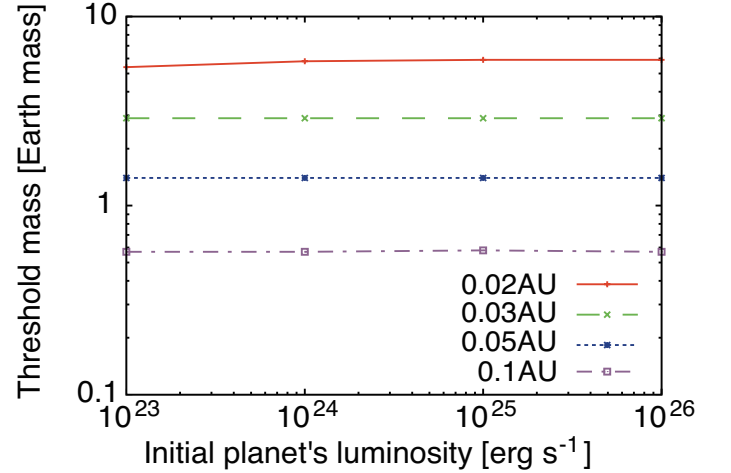}}
\caption{
The threshold mass in $M_{\oplus}$ as a function of the initial planet's luminosity in erg~s$^{-1}$ for four choices of semi-major axes.
The solid (red), dashed (green), dotted (blue), and dot-dashed (purple) represent $a=0.02, 0.03, 0.05$, and 0.1~AU, respectively.
We have assumed $X_{\mathrm{wt}}=$75~\% and $\varepsilon=0.1$.
}
\label{L_Mthrs}
\end{center}
\end{figure}

\subsection{Dependence on the initial water mass fraction} \label{MLRESLT3}
The fate of a water-rich planet also depends on the initial water mass fraction, $X_{\rm{wt},0}$.
Figure \ref{M_frac_wt_water} shows $X_{\rm{wt}}(t)$ at $t=10$~Gyr
as a function of the initial planet's mass, $M_{p,0}$, for four different values of $X_{\rm{wt},0}(=$
25~\%, 50~\%, 75~\%, and 100~\%).
As $M_{p,0}$ decreases, $X_{\rm{wt}}$(10~Gyr) decreases.
The pure water planet (solid line) with $M_{p,0}<0.82~M_{\oplus}$ is completely evaporated in 10~Gyr; namely, $X_{\rm{wt}}(10~\rm{Gyr})=$0~\%.
Otherwise, $X_{\rm{wt}}(10~\rm{Gyr})=100$~\%.
In other cases, we find that the threshold mass, $M_{\rm{thrs}}$, below which $X_{\rm{wt}}(10~\rm{Gyr})=$0~\%,
is
$0.56~M_{\oplus}$ for $X_{\rm{wt},0}=75$~\%, $0.44~M_{\oplus}$ for $X_{\rm{wt},0}=50$~\%, and $0.44~M_{\oplus}$ for $X_{\rm{wt},0}=25$~\%.
\begin{figure}[htbp]
\begin{center}
\resizebox{\hsize}{!}{\includegraphics{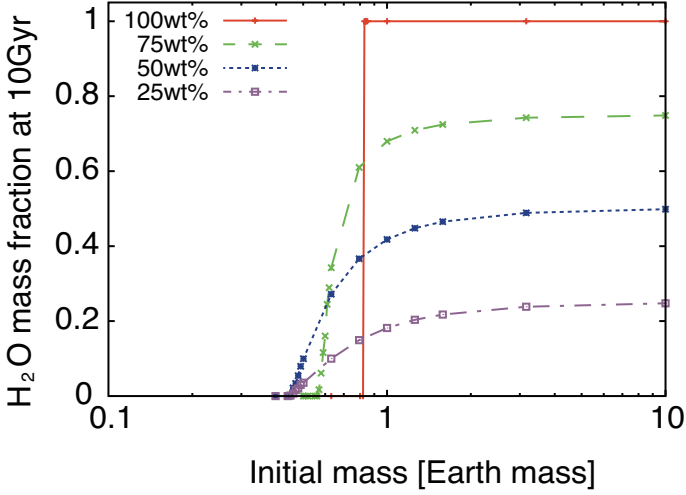}}
\caption{Relationship between the initial planetary mass and the fraction of the water envelope at 10~Gyr for four initial water mass fractions of  $X_{\rm{wt},0}=$ 100~\% (solid, red), 75~\% (dashed, green), 50~\% (dotted, blue), and 25~\% (dot-dashed, purple). We have assumed $L_{0}=1\times 10^{24}~\rm{erg~s}^{-1}$, $a=0.1$~AU, and $\varepsilon=0.1$. }
\label{M_frac_wt_water}
\end{center}
\end{figure} 

Figure \ref{M_frac_au_Xwt_water} shows the relationship between $X_{\rm{wt},0}$ and $M_{\rm{thrs}}$ for four different semi-major axes.
$M_{\rm{thrs}}$ is found not to be a monotonous function of $X_{\rm{wt},0}$. 
For $X_{\rm{wt},0}<25$~\%, $M_{\rm{thrs}}$ decreases, as $X_{\rm{wt},0}$ increases.
This is explained as follows.
According to Eq.~(\ref{result_tm_01}), the mass loss timescale {, $\tau_M$,} depends on the  {absolute amount of} water,
$X_{\rm{wt}}M_{p}$, and the planetary bulk density, $\rho_{\rm{pl}}$.
When $X_\mathrm{wt}$ is sufficiently small, $\rho_{\rm{pl}}$ is equal to the rocky density and is therefore constant. 
Thus, $\tau_M$ is determined only by the absolute amount of water (i.e., $X_\mathrm{wt} M_p$).
This means that, $M_p$ must be larger for $\tau_M$ to be the same if $X_\mathrm{wt,0}$ is small.
As a consequence, $M_\mathrm{thrs}$ decreases with increasing $X_\mathrm{wt,0}$.
More exactly, $M_{\rm{thrs}}$ changes with $X_{\rm{wt},0}$ in such a way that $X_{\rm{wt},0}M_{\rm{thrs}}$ is constant.
In contrast, when $X_\mathrm{wt,0}$ is large, $X_\mathrm{wt}$, $M_p$, and $\rho_\mathrm{pl}$ affect the mass loss timescale.
For a given $M_{p}$,
an increase in $X_{\rm{wt},0}$ leads to a decrease in $\rho_{\rm{pl}}$ (or, an increase in radius), which enhances mass loss.
As a result, $M_{\rm{thrs}}$ increases with $X_{\rm{wt},0}$ for $X_{\rm{wt},0}>25$~\%.
Therefore, there is a minimum value of $M_{\rm{thrs}}$, which is hereafter denoted by $M_{\rm{thrs}}^{\ast}$.

Similar trends can be seen in Figs.~3 and 4 of \citet{Lopez2013}.
To compare our results for water-rich planets to those for hydrogen-rich rocky planets from \citet{Lopez2013} in a more straightforward way, we show the relationship between the initial total mass and the fraction of the initial water envelope that is lost via subsequent photo-evaporation in 5~Gyr in Fig.~\ref{M_frac_wt_water2} (see Fig.~3c of \citealt{Lopez2013}). 
We set $L_{0}=1\times 10^{24}~\rm{erg~s}^{-1}$, $a=0.1$~AU, $\varepsilon=0.1$, and six initial water mass fractions of $X_{\rm{wt},0}=$ 1~\% (solid, red), 3~\% (long-dashed, green), 10~\% (dotted, blue), 30~\% (dash-dotted, purple), 50~\% (dot-dashed, light blue), and 60~\% (dashed black), 
which are similar to those adopted by \citet{Lopez2013}.
As mentioned above,
the initial total mass needed in the H+He case is larger by a factor of $\sim$10 than that in the water case
for the same fraction of the initial envelope to survive photo-evaporation.
In addition,
the required initial total mass for  $X_{\rm{wt},0} < 10~\%$ becomes significantly large
in the water case.
This behavior is also found in the case of the hydrogen-rich planets for $X_{\rm{wt},0} = 1-3~\%$. 
However, the trend is less noticeable in the H+He case. 
This is because the density effect described above is effective even for small H+He fractions.

\begin{figure}[htbp]
\begin{center}
\resizebox{\hsize}{!}{\includegraphics{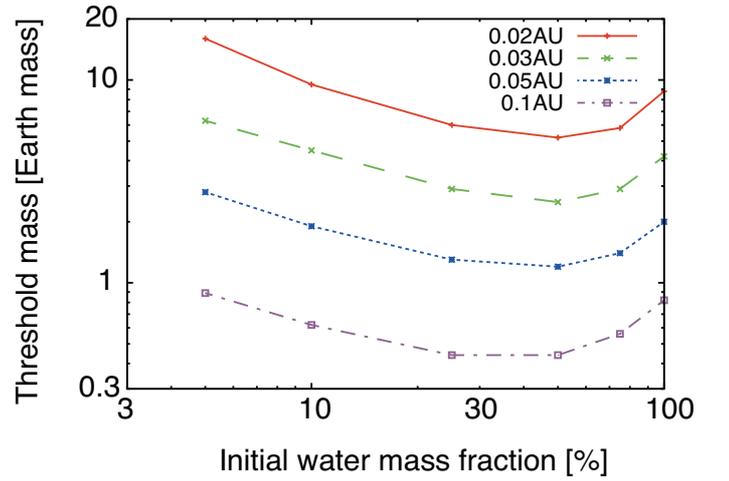}}
\caption{
Relationship between the initial water mass fraction $X_{\rm{wt},0}$ in \% and the threshold mass $M_{\rm{thrs}}$ in $M_{\oplus}$ for four choices of semi-major axes of 0.02~AU (solid, red), 0.03~AU (dashed, green), 0.05~AU (dotted, blue), and 0.1~AU (dot-dashed, purple).
We have assumed $L_{0}=1\times 10^{24}~\rm{erg~s}^{-1}$ and $\varepsilon=0.1$.
}
\label{M_frac_au_Xwt_water}
\end{center}
\end{figure}

\begin{figure}[htbp]
\begin{center}
\resizebox{\hsize}{!}{\includegraphics{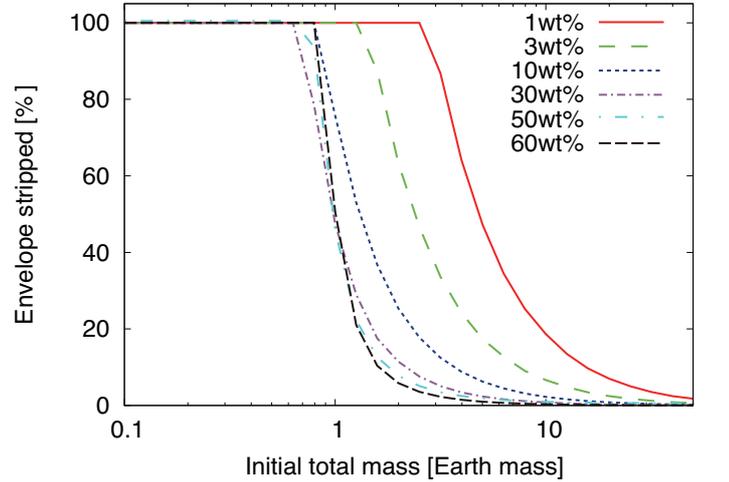}}
\caption{ {Relationship between the initial planetary mass and the fraction of the initial water envelope that is lost via photo-evaporation in 5~Gyr for six initial water mass fractions of  $X_{\rm{wt},0}=$ 1\% (solid, red), 3\% (long-dashed, green), 10\% (dotted, blue), 30\% (dash-dotted, purple), 50\% (dot-dashed, light blue), and 60\% (dashed black). We have assumed $L_{0}=1\times 10^{24}~\rm{erg~s}^{-1}$, $a=0.1$~AU, and $\varepsilon=0.1$. }}
\label{M_frac_wt_water2}
\end{center}
\end{figure} 

\subsection{Dependence on the semi-major axis} \label{MLRESLT4}
At small $a$, the incident stellar XUV flux becomes large.
Thus, $M_{\rm{thrs}}$ increases, as $a$ decreases.
Certainly, the distance to the host star
affects the equilibrium temperature $T_{\rm{eq}}$, which has an influence on $\rho_{\rm{pl}}$:
The higher $T_{\rm{eq}}$ is, the smaller $\rho_{\rm{pl}}$ is.
However, its impact on $M_{\rm{thrs}}$ is small, relative to that of $F_{\rm{XUV}}$.
According to the planet's mass and mean density relationship,
$\rho_{\rm{pl}}$ differs only by a factor of $\lesssim1.5$ between 880~K and 2000~K.
Therefore,
increasing $F_{\rm{XUV}}$ has a much greater impact on the mass loss than decreasing $\rho_{\rm{pl}}$.
In Fig.~\ref{M_frac_au_Xwt_water},
we find $M_{\rm{thrs}}^{\ast}=5.2~M_{\oplus}$ for $a=0.02$~AU, $M_{\rm{thrs}}^{\ast}=2.5~M_{\oplus}$ for $a=0.03$~AU, $M_{\rm{thrs}}^{\ast}=1.2~M_{\oplus}$ for $a=0.05$~AU, and $M_{\rm{thrs}}^{\ast}=0.44~M_{\oplus}$ for $a=0.1$~AU.


%
%

\subsection{Expected populations} \label{Mthrs_Rthrs}

Figure \ref{MR_w_obs} shows the relationship between $M_{\rm{thrs}}$ (not $M_{\rm{thrs}}^{\ast}$) and the radius
that the planet with $M_{\rm{thrs}}$ would have
at 10~Gyr without mass loss  {(solid line)}.
We call this radius the threshold radius, $R_{\rm{thrs}}$.
We have calculated $R_{\rm{thrs}}$ for $X_{\rm{wt},0}=$
100~\%, 75~\%, 50~\%, 25~\%, 10~\%, 5~\%, and 1~\%.
In addition, the mass-radius relationships for rocky planets  {(dashed line)} and pure-water planets  {(dotted line)} at 0.1~AU are also drawn in Fig.~\ref{MR_w_obs}.
There are four characteristic regions in Fig.~\ref{MR_w_obs}:
\begin{description}
\item[I] Planets must contain components less dense than water, such as hydrogen/helium.
\item[II] Planets with water envelopes and without H/He can exist. The water envelopes survive photo-evaporative mass loss.
\item[III] Primordial water  {envelopes experience significant} photo-evaporative mass loss in 10~Gyr.
\item[IV] Planets retain no water envelopes and are composed of rock and iron.
\end{description}
Only in the region~II, the planet retains
its primordial water envelope for 10~Gyr without significant loss.
There are minimum values not only of $M_{\rm{thrs}}$ but also of $R_{\rm{thrs}}$; the latter is denoted by $R_{\rm{thrs}}^{\ast}$ hereafter.
Note that $R_{\mathrm{thrs}}^{\ast}$ is not an initial radius. 

Those minimum values are helpful to discuss whether planets can possess water components or not,
because the uncertainty in water mass fractions can be removed.
Since $M_{\rm{thrs}}$ and $R_{\rm{thrs}}$
depend on semi-major axis, we also compare those threshold values with 
observed $M-a$ and $R-a$ relationships in the next section.
\begin{figure}[htbp]
\begin{center}
\resizebox{\hsize}{!}{\includegraphics{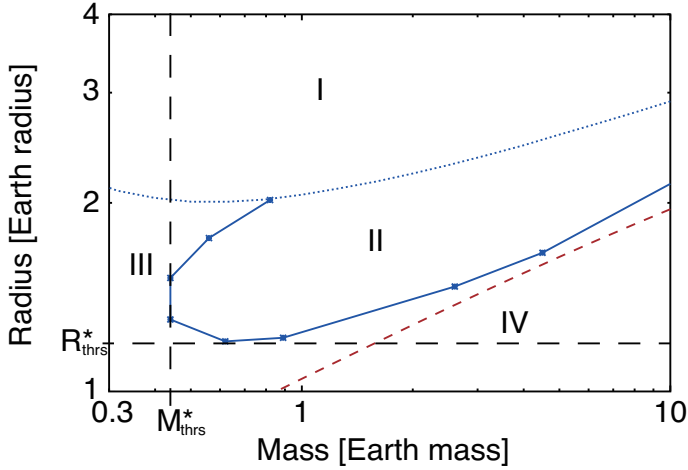}}
\caption{
Relationship  {between} the threshold mass and  {the threshold radius. 
The latter is defined by the radius} 
that the planet with $M_{\rm{thrs}}$ would have at 10~Gyr without  {ever experiencing} mass loss
( {denoted by} $R_{\rm{thrs}}$).
The squares, which are connected with a solid line, are $M_{\rm{thrs}}$ and $R_{\rm{thrs}}$ for 0.1~AU
and 
seven different initial water mass fractions $X_{\rm{wt,0}}=100~\%$, 75~\%, 50~\%, 25~\%, 10~\%, 5~\%, 1~\%, and 0.5~\%.
The dashed and dotted lines represent
mass-radius relationships, respectively, for
rocky planets and pure-water planets at 0.1~AU.
$M_{\rm{thrs}}^{\ast}$ and $R_{\rm{thrs}}^{\ast}$ represent the minimum values of $M_{\rm{thrs}}$ and $R_{\rm{thrs}}$, respectively.
}
\label{MR_w_obs}
\end{center}
\end{figure}

%
%

\section{Implications for distributions of observed exoplanets}\label{hikaku}

Figure~\ref{MR_thrs} compares the relationship between the threshold mass, $M_{\rm{thrs}}$, and threshold radius, $R_{\rm{thrs}}$ with measured masses and radii of super-Earths around G-type stars identified so far. 
Here we show three theoretical relationships for $a = 0.02$, $0.05$, and $0.1$~AU.
As discussed above,
only planets on the right side of the theoretical line (i.e., in region~II) for a given $a$ are able to retain their water envelopes without significant loss for 10~Gyr.

For future characterizations, planets in region~III would be of special interest, because our results suggest that planets should be rare in region~III.
Three out of the 14 planets, 55~Cnc~e, Kepler-20~b, and CoRoT-7~b
might be in region~III,
although errors and the uncertainty in $\varepsilon$ (see also the lower panel of Figure \ref{Mthrs_eps} for the sensitivity of $M_{\rm{thrs}}^{\ast}$ to $\varepsilon$) are too large to conclude so.
There are at least three possible scenarios for the origin of planets in region~III.
One is that those planets are halfway to complete evaporation of their water envelopes.
Namely, some initial conditions happen to make planets in region~III, although such conditions are rare.
The second  {possible scenario} is that those planets had formed far from and migrated toward their host stars recently.
The third is that those planets are in balance between degassing from the rocky core and the atmospheric escape.
Thus, deeper understanding of the properties of those super-Earths via future characterization will provide important constraints on their origins.
\begin{figure}[htbp]
\begin{center}
\resizebox{\hsize}{!}{\includegraphics{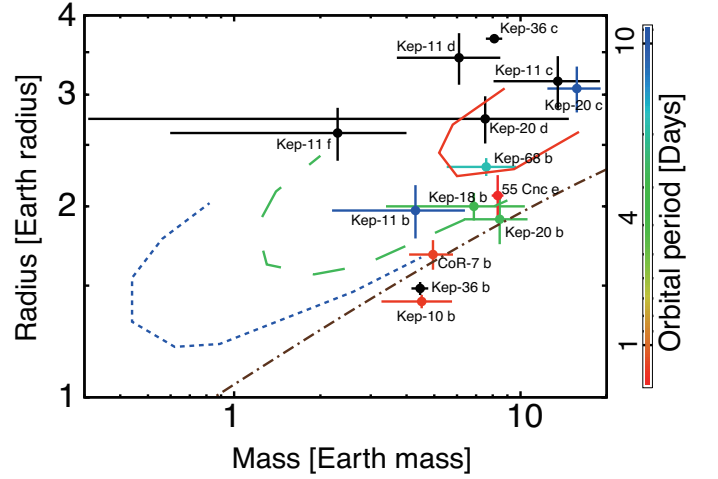}}
\caption{
Relationship between the threshold mass $M_{\rm{thrs}}$ and radius $R_{\rm{thrs}}$ (lines; see text for definitions)
compared to masses and radii of observed transiting super-Earths around G-type stars (points with error bars; exoplanets.org (\citealt{Wright2011}), as of June 29, 2013, ).
The dotted (blue), dashed (green), and solid (red) represent the $M_{\mathrm{thrs}}$ and $R_{\mathrm{thrs}}$ relationships for orbital periods of 11~days (=~0.1~AU), 4~days (=~0.05~AU), and 1~day (=~0.02~AU), respectively.
The dash-dotted (brown) line represents the planet composed of rocks.
Note that black points represent planets whose orbital periods are longer than 11days.
In those calculations, we have assumed the heating efficiency $\varepsilon=0.1$ and the initial luminosity $L_0=1\times 10^{24}~\rm{erg~s}^{-1}$.
"CoR" are short for CoRoT and "Kep" are short for Kepler.}
\label{MR_thrs}
\end{center}
\end{figure}

In this study, low-mass exoplanets, whose masses are $\le 20~M_{\oplus}$ and radii $\le 4~R_{\oplus}$, are of special interest. 
(We call them super-Earths below.) 
While there are only 14 super-Earths  
whose masses and radii were both measured (see Fig.~\ref{MR_thrs}),
the minimum masses ($M_p \sin {i}$) and the orbital periods were measured for about 22 super-Earths around G-type stars (see Fig.~\ref{Mthrs_eps}).
Also, over 1,000 sub/super-Earth-sized planet candidates have been identified by the Kepler space telescope (Batalha et al.~2013). The size and semi-major axis distribution of those objects is known. It is, thus, interesting to compare our theoretical prediction with the observed $M_p$-$a$ and $R_p$-$a$ distributions. 

Before doing so, we demonstrate that $M_{\rm{thrs}}^{\ast}$ and $R_{\rm{thrs}}^{\ast}$ are good indicators for constraining the limits below which \textit{evolved} planets retain no water envelopes. 
Figure~\ref{Mthrs_eps}a and \ref{Rthrs_eps}a show the theoretical distributions of masses and radii of planets that evolved for 10~Gyr, starting with various initial water mass fractions and planetary masses (i.e., $X_{\rm{wt},0}=25, 50, 75$ and 100~\% and $\log(M_{p,0}/M_{\oplus})=-1+0.1j$ with $j=0, 1, \cdots, 21)$.
The crosses (red) and open squares (blue) represent the planets that lost their water envelopes completely (i.e., rocky planets) and those which survive significant loss of their water envelopes, respectively.
As seen in these figures, two populations of rocky planets and water-rich planets are clearly separated by the $M_{\rm{thrs}}^{\ast}$ and $R_{\rm{thrs}}^{\ast}$ lines.
Note that there are some planets that retain their water envelope below the threshold line.
These planets just retain $\lesssim$ 1~\% water mass fraction at 10~Gyr.
However, such planets are found to be obviously rare.

In Fig.~\ref{Mthrs_eps}b, we show the distribution of $M_{p}\sin i$ and $a$ of low-mass exoplanets detected around G-type and K-type stars so far, as compared with $M_{\rm{thrs}}^{\ast}$ for three choices of $\varepsilon$.
Among them, $\alpha$~Cen B~b, Kepler-10~b and CoRoT-7~b are well below the $M_\mathrm{thrs}^\ast$ line for $\varepsilon = 0.1$.
Thus, the three planets are likely to be rocky, provided $\varepsilon=0.1$.
However, the uncertainty in $\varepsilon$ (and $F_\mathrm{XUV}$) prevents us from deriving a robust conclusion.
An order-of-magnitude difference in $\varepsilon$ is found to change $M_\mathrm{thrs}^\ast$ by a factor of three. 
The aforementioned three planets are between the two $M_\mathrm{thrs}^{\ast}$ lines for $\varepsilon =$ 0.01 and 0.1. This demonstrates quantitatively how important determining $\varepsilon$ and $F_\mathrm{XUV}$ more accurately is for understanding the composition of super-Earths only with measured masses. 
It would be worth mentioning that few planets are found between the lines for $\varepsilon =$ 0.1 and $\varepsilon =$ 1. 
Since all the planets in Fig.~\ref{Mthrs_eps}b were found by the radial-velocity method, 
the apparent gap would be unlikely to be due to observational bias. 
Thus, the gap might suggest that the actual $M_\mathrm{thrs}^\ast$ line lies between those two ones. 

In Fig.~\ref{Rthrs_eps}b, we show the distribution of $R_p$ and $a$ of KOIs, which is compared with $R_\mathrm{thrs}^{\ast}$
for three choices of $\varepsilon$.
Many planets are found to be below the $R_{\rm{thrs}}^{\ast}$ lines.
We are unable to constrain the fraction of rocky planets quantitatively, because of the uncertainty in $\varepsilon$. 
However, since there are many points below the $R_{\rm{thrs}}^{\ast}$ line for $\varepsilon$ of as small as 0.01, it seems to be a robust conclusion that KOIs contain a significant number of rocky planets.
Note that the distribution must include rocky planets that were formed rocky without ever experiencing mass loss.
This means that there are more rocky planets in reality than we have predicted in this study.

As mentioned in Introduction, \citet{Lopez2013} performed a similar investigation of the threshold mass and radius concerning H+He atmospheres on rocky super-Earths (see Figs.~8 and 9 of \citet{Lopez2013}). 
For the horizontal axis, they adopted the incident stellar flux, instead of semi-major axis. 
In Figs.~\ref{Mthrs_eps} and \ref{Rthrs_eps}, we have also indicated another scale of the incident flux calculated from the relationship between the semi-major axis $a$ and the incident flux $F$,  
\begin{equation}
F = \frac{L_{\mathrm{star}}}{4\pi a^2} = F_{\mathrm{Earth}}\left( \frac{L_{\mathrm{star}}}{L_{\odot}} \right) \left( \frac{a}{1\mathrm{AU}}\right)^{-2},
\end{equation}
where $L_{\mathrm{star}}$ is the luminosity of the host star and $F_{\mathrm{Earth}}$ is the current bolometric flux that the Earth receives from the Sun.
Comparing their results for the H+He envelope, 
we find that
the threshold value of the initial mass (or incident flux) for H$_2$O is smaller by a factor of about 10 than that for H+He
although a similar linear dependence is found.
For example,
the threshold mass for H+He is $\sim 30M_{\oplus}$ (derived by Eq.~(6) of \citealt{Lopez2013}) in the case of $F=10^{3}F_{\oplus}$,
while it is for H$_2$O is $\sim 2M_{\oplus}$. 

In Fig.~9 of \citet{Lopez2013}, it has also been suggested that the frequency of planets with radii of $1.8-4.0~R_{\oplus}$ for $F_{p}\ge 100~F_{\oplus}$ (corresponding to $a \le 0.1~\textrm{AU}$) should be low as a consequence of photo-evaporative mass loss. 
\citet{Owen2013} also found a deficit of planets around $2~R_{\oplus}$ in their planet distribution (see Fig.~8 of \citealt{Owen2013}). In contrast, our results suggest that water-rich planets with radii of $1.5-3.0~R_{\oplus}$ are relatively common, because they are able to sustain their water envelopes against photo-evaporation. 
This seeming disagreement on the predicted distribution demonstrates the influence of the envelope composition
on the predicted distribution.
Indeed, there are many KOIs found in such a domain in the $R_p$-$a$ diagram shown in Fig.~\ref{Rthrs_eps}a. 
Thus, those KOIs may be water-rich planets, although it is also possible that they are rocky planets without ever experiencing mass loss.

Finally, we focused in this study on the thermal escape of the upper atmosphere due to stellar XUV irradiation. In addition, ion pick-up induced by stellar winds and coronal mass ejections may be effective in stripping off atmospheres of close-in planets, as discussed for close-in planets with hydrogen-rich atmospheres (e.g. \citealt{Lammer2013}).
Such non-thermal effects lead to increase in $M_{\rm{thrs}}^{\ast}$.
This implies that the $M_{\rm{thrs}}^{\ast}$ obtained in this study is a lower limit on survival of water-rich planets.

\begin{figure}[htbp]
\begin{center}
\resizebox{\hsize}{!}{\includegraphics{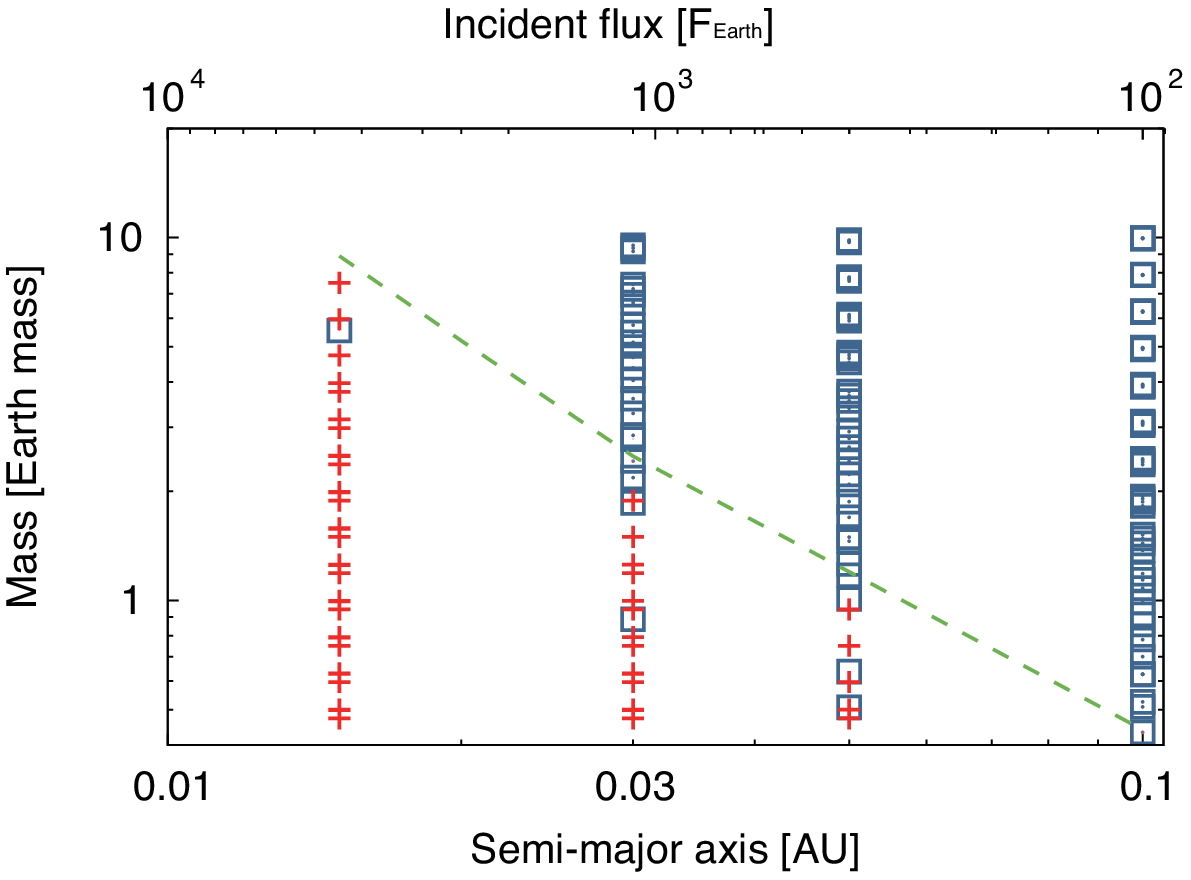}}
\resizebox{\hsize}{!}{\includegraphics{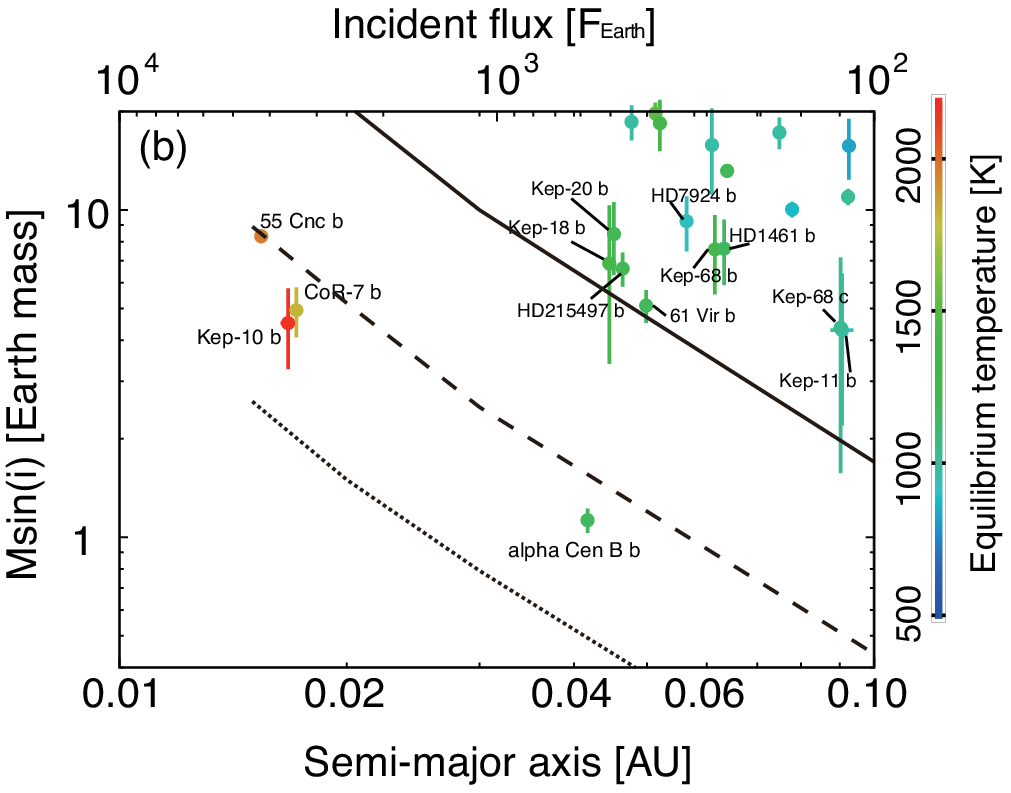}}
\caption{
\textit{Upper panel}:(a) Theoretical distribution of masses and semi-major axes  {(or incident fluxes)} of planets at 10~Gyr with various initial masses and water mass fractions.
Cross (red) points represent planets that lost their water envelopes completely in 10~Gyr, while open squares (blue) represent planets that survive significant loss of the water envelopes via photo-evaporation.
The green line is the minimum threshold masses, $M_{\rm{thrs}}^{\ast}$.
Here, we have adopted $\varepsilon=0.1$.
\textit{Lower panel}:(b) Distribution of masses and semimajor axes   {(or incident fluxes)} of detected exoplanets compared to the minimum threshold mass, $M_{\rm{thrs}}^{\ast}$, derived in this study (see section~3.3 for definition).
We have shown three $M_{\rm{thrs}}^{\ast}-a$ relationships for different heating efficiencies: $\varepsilon=1$ (solid line), $\varepsilon=0.1$ (dashed line), and $\varepsilon=0.01$ (dotted line).
Filled circles with error bars
represent observational data (from http://exoplanet.org~(\citealt{Wright2011}))
for planets orbiting host stars with effective temperature of 5000-6000~K (relatively early K-type stars and G-type stars).
Planets are colored according to
their zero-albedo equilibrium temperatures in K.
In the planet names, "CoR" and "Kep" stand for CoRoT and Kepler, respectively.
}
\label{Mthrs_eps}
\end{center}
\end{figure}

\begin{figure}[htbp]
\begin{center}
\resizebox{\hsize}{!}{\includegraphics{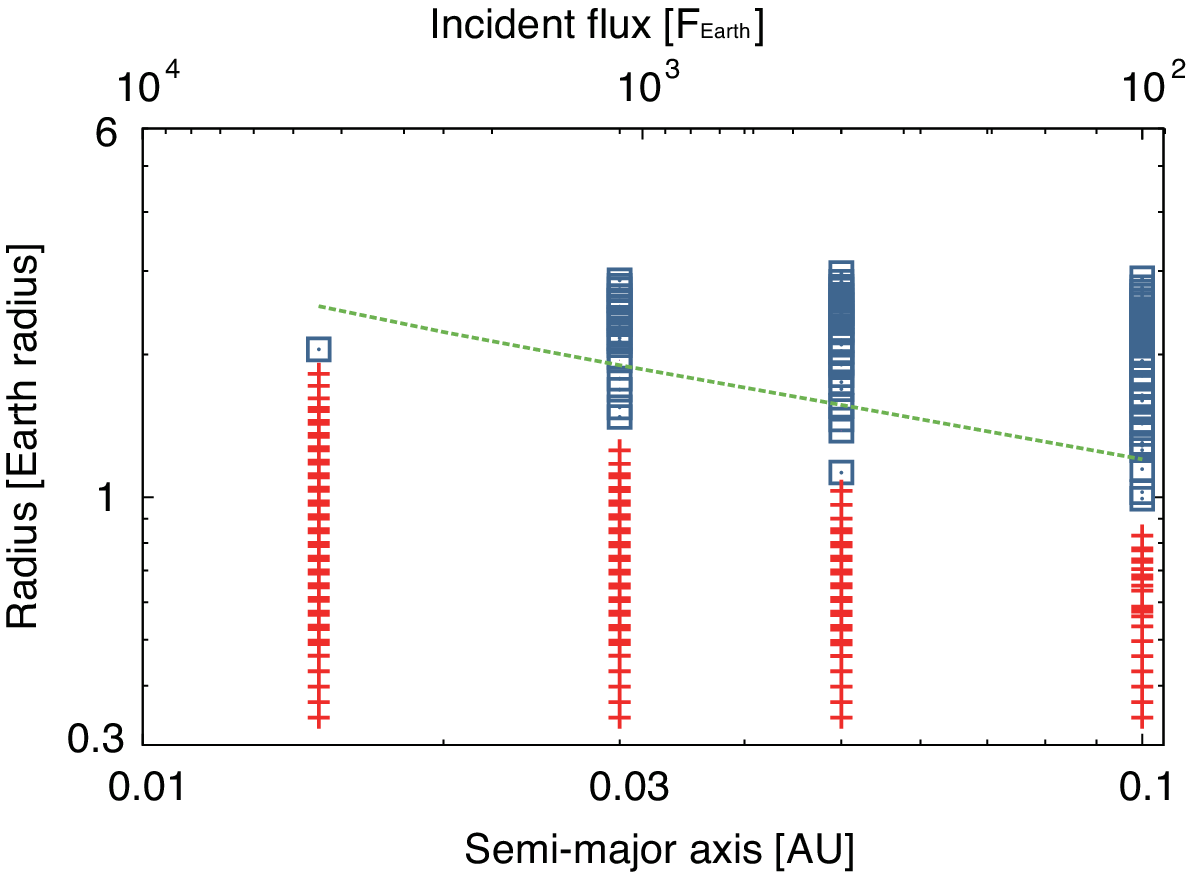}}
\resizebox{\hsize}{!}{\includegraphics{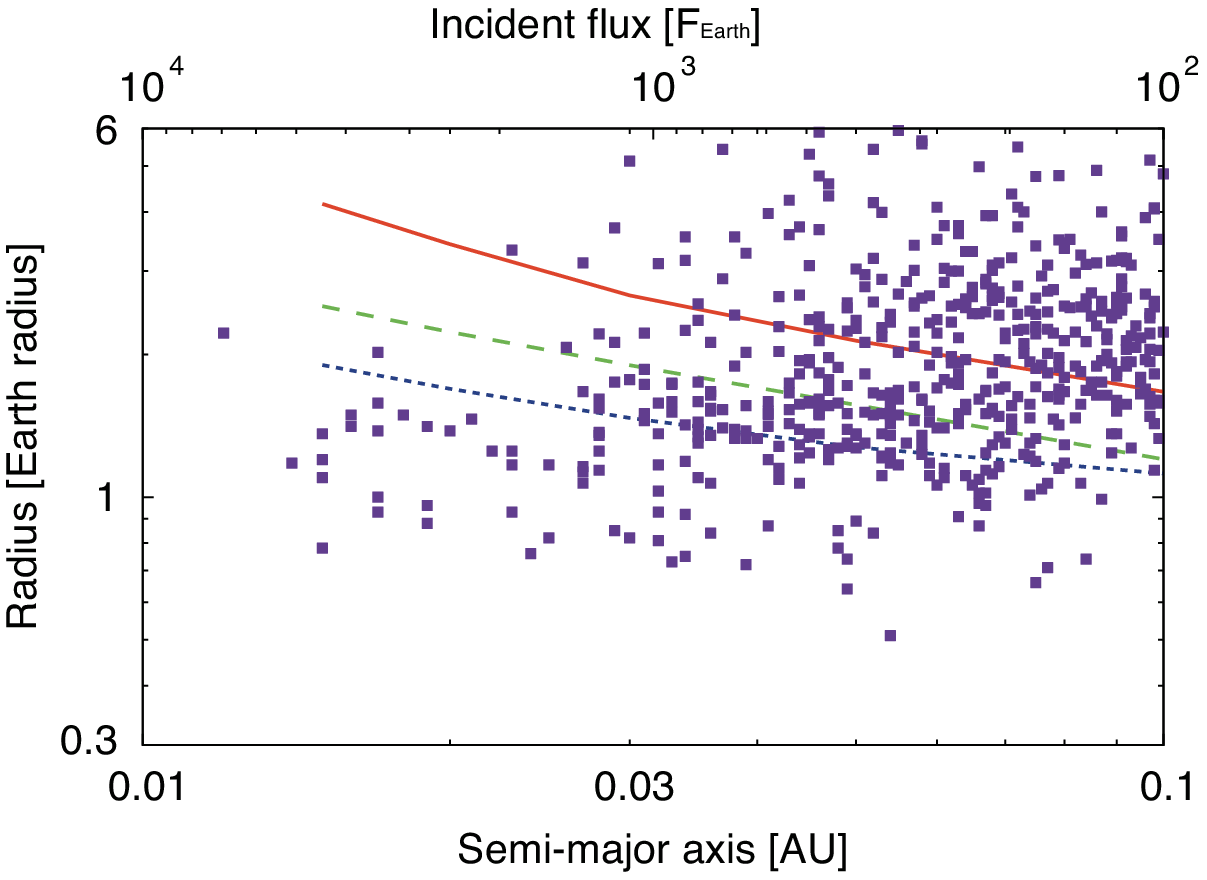}}
\caption{
\textit{Upper panel}:(a) Theoretical distribution of radii and semi-major axes  {(or incident fluxes)} of planets at 10~Gyr with various initial masses and water mass fractions.
Cross (red) points represent planets that lost their water envelopes completely due to the photo-evaporation in 10~Gyr, while open squares (blue) represent planets that survive significant loss of the water envelopes.
The green line is the minimum threshold radii, $R_{\rm{thrs}}^{\ast}$.
Here, we have adopted $\varepsilon=0.1$. 
\textit{Lower panel}:(b) Distribution of radii and semi-major axes  {(or incident fluxes)} of Kepler planetary candidates, compared to the threshold radius, $R_{\rm{thrs}}^{\ast}$ (see section~3.3 for definition).
We have shown three $R_{\rm{thrs}}^{\ast}-a$ relationships for different heating efficiencies: $\varepsilon=1$ (red solid line), $\varepsilon=0.1$ (green dashed line),
and $\varepsilon=0.01$ (blue dotted line).
Filled squares represent observational data (http://kepler.nasa.gov, as of June 29, 2013) for planets orbiting host stars with effective temperature of 5300-6000~K (G-type stars).
}
\label{Rthrs_eps}
\end{center}
\end{figure}

%
%
\section{Summary}

In this study, we have investigated the impact of photo-evaporative mass loss on masses and radii of water-rich sub/super-Earths with short orbital periods around G-type stars.
We simulated the interior structure and the evolution of highly-irradiated sub/super-Earths that consist of a rocky core surrounded by a water envelope, including the effect of mass loss due to the stellar XUV-driven energy-limited hydrodynamic escape (see section~\ref{model}).

The findings from this study are summarized as follows.
In section~\ref{MLresult}, we have investigated the mass evolution of water-rich sub/super-Earths, 
and then found a threshold planet mass $M_{\rm{thrs}}$,
below which the planet has its water envelope stripped off in 1-10~Gyr (section~\ref{MLRESLT1}).
The initial planet's luminosity has little impact on $M_{\rm{thrs}}$ (section~\ref{MLRESLT2}).
We have found that there is a minimum value, $M_{\rm{thrs}}^{\ast}$, for given $a$ and $\varepsilon$ (section~\ref{MLRESLT4}). Water-rich planets with initial masses smaller than $M_{\rm{thrs}}^{\ast}$ lose their water envelopes completely in 10~Gyr, independently of initial water mass fraction.
The threshold radius, $R_{\rm{thrs}}$, is defined as the radius that the planet of mass $M_{\rm{thrs}}$ would have at 10~Gyr if it evolved without undergoing mass loss. 
We have also found that there is a minimum value of the threshold radius, $R_{\rm{thrs}}^{\ast}$
(section~\ref{Mthrs_Rthrs}).
Finally, we have discussed the composition of observed exoplanets in section~\ref{hikaku}
by comparing the threshold values to measured masses and radii of the exoplanets.
Then, we have confirmed quantitatively that more accurate determination of planet masses and radii, $\epsilon$ and $F_{\rm{XUV}}$, respectively is needed for deriving robust prediction for planetary composition.
Nevertheless, the comparison between $R_{\rm{thrs}}^{\ast}$ and radii of KOIs in the $R_p-a$ plane suggests that KOIs contain a significant number of rocky planets.

In this study, we have demonstrated that photo-evaporative mass loss has a significant impact on the evolution of water envelopes of sub/super-Earths, especially with short orbital periods,
and that of H+He envelopes of super-Earths.
Since the $M_{\mathrm{thrs}}$ for water envelope models is larger by a factor of 10, relative to that for H+He envelope models by \citet{Lopez2013},
the stability limit for water envelopes gives more robust constraints on the detectability of rocky planets.
Thus, the $M_{\rm{thrs}}$ and $R_{\rm{thrs}}$ will provide valuable information for future searches of rocky Earth-like planets.

%
%
\section*{Acknowledgements}
We thank N. Nettelmann for providing us with tabulated data for equation of state of water ( H$_{2}$O-EOS) and S. Ida and T. Guillot for fruitful advices and discussions.  {We also thank the anonymous referee for his/her careful reading and constructive comments that helped us improve this paper greatly.} 
 {We also thank Y. Ito and Y. Kawashima for providing us with the opacity data and fruitful suggestions about the atmospheric structure.}
 {This research has made use of the Exoplanet Orbit Database and the Exoplanet Data Explorer at exoplanets.org.}
This study is supported by Grants-in-Aid for Scientific Research on Innovative Areas (No. 23103005) and Scientific Research (C) (No. 25400224) from the Ministry of Education, Culture, Sports, Science and Technology (MEXT) of Japan.
K. K. is supported by a grant for the Global COE Program, ''From the Earth to ''Earths'''', of MEXT, Japan.
Y. H. is supported by the Grant-in-Aid for JSPS Fellows (No. 23003491) from MEXT, Japan.

\bibliography{kms_v12+ikoma_3}	

\appendix
\section{Atmospheric model}

First, we describe opacity models for the water vapor atmosphere.
We define the Planck-type ($\kappa^\mathrm{P}$) and the Rosseland-type mean opacities ($\kappa^\mathrm{r}$) as
\begin{eqnarray}
 \kappa_{\mathrm{v}}^{\mathrm{p}} &=& \int_\mathrm{visible} \kappa_{\nu} B_{\nu}(T_{\star})d\nu~\bigg/ \int_\mathrm{visible} B_{\nu}(T_{\star})d\nu, \\
\frac{1}{ \kappa_{\mathrm{v}}^{\mathrm{r}}} &=& \int_\mathrm{visible} \frac{1}{\kappa_{\nu}}\frac{dB_{\nu}(T_{\star})}{dT}d\nu ~\bigg/ \int_\mathrm{visible} \frac{dB_{\nu}(T_{\star})}{dT}d\nu \label{kapvr}, \\
 \kappa_{\mathrm{th}}^{\mathrm{p}} &=& \int_\mathrm{thermal} \kappa_{\nu} B_{\nu}(T_{\mathrm{atm}})d\nu~\bigg/ \int_\mathrm{thermal} B_{\nu}(T_{\mathrm{atm}})d\nu, \\
 \frac{1}{\kappa_{\mathrm{th}}^{\mathrm{r}}} &=& \int_\mathrm{thermal} \frac{1}{\kappa_{\nu}}\frac{dB_{\nu}(T_{\mathrm{atm}})}{dT}d\nu~\bigg/ \int_\mathrm{thermal} \frac{dB_{\nu}(T_{\mathrm{atm}})}{dT}d\nu,
\end{eqnarray}
where $\nu$ is the frequency; $\kappa_\nu$ the monochromatic opacity at a given $\nu$; $T_{\star}$ the stellar effective temperature; $T_{\mathrm{atm}}$ the atmospheric temperature of the planet; and $B_\nu$ the Planck function. The subscripts, "th" and "v", mean opacities in the thermal and visible wavelengths, respectively.
In this study, we assume $T_{\star}=$5780~K.
We adopt HITRAN opacity data for water \citep{Rothman2009} and calculate mean opacities for 1000~K, 2000~K, and 3000~K at 1, 10, 100~bar. Mean opacities are fitted to power-law functions of $P$ and $T$, using the least squares method;
\begin{eqnarray}
\kappa_{\mathrm{v}}^{\mathrm{p}} &=& 1.94\times10^4 \left(\frac{P}{1\mathrm{bar}} \right)^{0.01} \left( \frac{T}{1000\mathrm{K}} \right)^{1.0} \mathrm{cm}^2~\mathrm{g}^{-1}, \label{kap1} \\
\kappa_{\mathrm{v}}^{\mathrm{r}} &=& 2.20 \left(\frac{P}{1\mathrm{bar}} \right)^{1.0} \left( \frac{T}{1000\mathrm{K}} \right)^{-0.4} \mathrm{cm}^2~\mathrm{g}^{-1}, \label{kap2} \\
\kappa_{\mathrm{th}}^{\mathrm{p}} &=& 4.15\times10^{5} \left(\frac{P}{1\mathrm{bar}} \right)^{0.01} \left(\frac{T}{1000\mathrm{K}} \right)^{-1.1} \mathrm{cm}^2~\mathrm{g}^{-1}, \label{kap3} \\
\kappa_{\mathrm{th}}^{\mathrm{r}} &=&3.07 \times 10^{2} \left(\frac{P}{1\mathrm{bar}} \right)^{0.9} \left(\frac{T}{1000\mathrm{K}} \right)^{-4.0} \mathrm{cm}^2~\mathrm{g}^{-1}, \label{kap4}
\end{eqnarray}
where $P$ is the pressure and $T$ the temperature.

In this study, we basically follow the prescription developed by \citet{Guillot2010} except for the treatment of the opacity.
We consider a static, plane-parallel atmosphere in local thermodynamic equilibrium. We assume that the atmosphere is in radiative equilibrium between an incoming visible flux from the star and an outgoing infrared flux from the planet. Thus, the radiation energy equation and radiation momentum equation are written as  
\begin{eqnarray}
\frac{dH_{\mathrm{v}}}{dm} &=& \kappa_{\mathrm{v}}^{\mathrm{p}}J_{\mathrm{v}}, \label{RMT-01}\\
\frac{dK_{\mathrm{v}}}{dm} &=& \kappa_{\mathrm{v}}^{\mathrm{r}}H_{\mathrm{v}}, \label{RMT-02} \\
\frac{dH_{\mathrm{th}}}{dm} &=& \kappa_{\mathrm{th}}^{\mathrm{p}}\left( J_{\mathrm{th}} - B \right), \label{RMT-03}\\
\frac{dK_{\mathrm{th}}}{dm} &=& \kappa_{\mathrm{th}}^{\mathrm{r}}H_{\mathrm{th}}, \label{RMT-04}
\end{eqnarray}
and the atmosphere in radiative equilibrium satisfies
\begin{eqnarray}
\kappa_{\mathrm{v}}^{\mathrm{p}}J_{\mathrm{v}} + \kappa_{\mathrm{th}}^{\mathrm{p}}\left( J_{\mathrm{th}} -B\right) = 0, \label{RMT-05}
\end{eqnarray}
where $J_\mathrm{v}$ ($J_\mathrm{th}$), $H_\mathrm{v}$ ($H_\mathrm{th}$), and $K_\mathrm{v}$ ($K_\mathrm{th}$) are, respectively, the zeroth-, first-, and second-order moments of radiation intensity in the visible (thermal) wavelengths, $m$ the atmospheric mass coordinate, $dm = \rho dz$, where $z$ is the altitude from the bottom of the atmosphere, $\rho$ the density, and $B$ the frequency-integrated Planck function,
\begin{equation}
B \equiv \int_\mathrm{thermal} B_\nu d\nu \sim \frac{\sigma}{\pi}T^4,
\end{equation}
where $\sigma$ is the Stefan-Boltzmann constant.
We assume here that thermal emission from the atmosphere at visible wavelengths are negligible, so that $B_\nu \sim 0$ in the visible region.
The six moments of the radiation field are defined as 
\begin{eqnarray}
(J_\mathrm{v}, H_\mathrm{v}, K_\mathrm{v}) &\equiv& \int_\mathrm{visible} (J_\nu, H_\nu, K_\nu) d\nu, \\
(J_\mathrm{th}, H_\mathrm{th}, K_\mathrm{th}) &\equiv& \int_\mathrm{thermal} (J_\nu, H_\nu, K_\nu) d\nu,
\end{eqnarray}
where $J_\nu$ is the mean intensity, $4\pi H_\nu$ the radiation flux, and $4\pi K_\nu/c$ the radiation pressure ($c$ is the speed of light).

We integrate three moments of specific intensity, $J_\nu,H_\nu$ and $K_\nu$, over all the frequencies:
\begin{eqnarray}
J &\equiv& \int_0^{\infty} J_\nu d\nu = \frac{1}{2} \int_0^{\infty} d\nu \int_{-1}^{1}  d\mu I_{\nu,\mu} = J_{\mathrm{v}} + J_{\mathrm{th}}, \\
H &\equiv& \int_0^{\infty} H_\nu d\nu = \frac{1}{2} \int_0^{\infty} d\nu \int_{-1}^{1}  d\mu I_{\nu,\mu}\mu = H_{\mathrm{v}} + H_{\mathrm{th}}, \\
K &\equiv& \int_0^{\infty} K_\nu d\nu = \frac{1}{2} \int_0^{\infty} d\nu \int_{-1}^{1}  d\mu I_{\nu,\mu}\mu^2 = K_{\mathrm{v}} + K_{\mathrm{th}},
\end{eqnarray}
where $I_{\nu,\mu}$ is the specific intensity and $\theta$ the angle of a intensity with respect to the $z$-axis, $\mu=\cos\theta$. The energy conservation of the total flux implies 
\begin{equation}
H = H_{\mathrm{v}} + H_{\mathrm{th}} = \frac{1}{4\pi} \sigma T_{\mathrm{int}}^4, \label{AT01a}
\end{equation}
where $T_{\mathrm{irr}}$ is the irradiation temperature given by
\begin{equation}
T_{\mathrm{irr}} = T_{\star}\sqrt{\frac{R_{\star}}{a}},
\end{equation}
where $R_{\star}$ is the radius of the host star and $a$ the semi-major axis.

For the closure relations, we use the Eddington approximation (e.g. \citealt{Chandrasekhar1960}), namely,
\begin{eqnarray}
K_{\mathrm{v}} &=& \frac{1}{3}J_{\mathrm{v}}, \\
K_{\mathrm{th}} &=& \frac{1}{3}J_{\mathrm{th}}.
\end{eqnarray}
For an isotropic case of both the incoming and outgoing radiation fields, we find boundary conditions of the moment equations as follows (see also \citealt{Guillot2010} for details):
\begin{eqnarray}
H_{\mathrm{v}}(m=0) &=& -\frac{1}{\sqrt{3}}\frac{1}{4\pi} \sigma T_{\mathrm{irr}}^4, \label{AT01b}\\
H_{\mathrm{v}}(m=0) &=& -\frac{1}{\sqrt{3}} J_{\mathrm{v}}(m=0), \label{AT01c}\\
H_{\mathrm{th}}(m=0) &=& \frac{1}{2} J_{\mathrm{th}}(m=0). \label{AT01d}
\end{eqnarray}

Thus, we integrate Eqs.(\ref{RMT-01})-(\ref{RMT-05}) over $m$ numerically, using mean opacities of (\ref{kap1})-(\ref{kap4}) and boundary conditions of (\ref{AT01b})-(\ref{AT01d}), and then determine a T-P profile of the water vapor atmosphere.
We assume that the boundary is at $P_{0}=1\times 10^{-5}$~bar.
The choice of $P_0$ ($\le 1\times 10^{-5}$bar) has little effect on the atmospheric temperature-pressure structure. 
$T_0$ is determined in an iterative fashion
until $\mathrm{abs}( T_{0}-[\pi B(m=0, P_0, T_0)/\sigma]^{1/4} ) \le 0.01$
is fulfilled.
Then we integrate Eqs.~(\ref{RMT-01})-(\ref{RMT-05}) over $m$ by the 4th-order Runge-Kutta method,
until we find the point where $\mathrm{d}\ln T/ \mathrm{d}\ln P \ge \nabla_{\mathrm{ad}}$. 
The pressure and temperature, $P_{\mathrm{ad}}$ and $T_{\mathrm{ad}}$, are the boundary conditions for the convective-interior structure (see section\ref{equations}).

In Fig.~\ref{atm_hikaku}, we show the $P$-$T$ profile for the solar-composition atmosphere with $g=980$~cm~s$^{-2}$, $T_{\mathrm{int}}=300$~K, and $T_{\mathrm{irr}}=1500$~K (dotted line). 
In this calculation, we take $\kappa_{\mathrm{th}}^{\mathrm{r}}$ and $\kappa_{\mathrm{th}}^{\mathrm{p}}$ as functions of $P$ and $T$ from \citet{Freedman2008} and calculate 
$\kappa_\mathrm{v}^\mathrm{p}$ and $\kappa_\mathrm{v}^\mathrm{r}$, for $P=1\times10^{-3}, 0.1,1, 10$~bar and $T=1500$~K from HITRAN and HITEMP data that include
H$_2$, He, H$_2$O, CO, CH$_{4}$, Na, and K for the solar abundance respectively as
\begin{eqnarray}
\kappa_{\mathrm{v}} = \left\{  
\begin{array}{llclc}
1.51\times 10^{-5} & \mathrm{cm}^2 \cdot \mathrm{g}^{-1} & (10^{-3} &\le P[\mathrm{bar}]),& \\
3.88\times 10^{-4} & \mathrm{cm}^2 \cdot \mathrm{g}^{-1} & (10^{-3} &< P[\mathrm{bar}] \le& 10^{-1}), \\
3.05\times 10^{-3} & \mathrm{cm}^2 \cdot \mathrm{g}^{-1} & (10^{-1} &< P[\mathrm{bar}] \le& 1),  \\
2.65\times 10^{-2} & \mathrm{cm}^2 \cdot \mathrm{g}^{-1} & (1  &> P[\mathrm{bar}]), & \\
\end{array}
\right.
\end{eqnarray}
by use of (\ref{kapvr}).
The thin and thick parts of the dotted line represent the radiative and convective zones, respectively.

In addition, we test our atmosphere model by comparing it with the $P$-$T$ profile derived by \citet{Guillot2010} with $\gamma=\kappa_{\mathrm{v}}/\kappa_{\mathrm{th}}=0.4$ (solid line), which reproduces more detailed atmosphere models by \citet{Fortney2005} and \citet{Iro2005} (see Fig.~6 of \citet{Guillot2010}).
As seen in Fig.~\ref{atm_hikaku}, our atmospheric model yields a $P$-$T$ profile similar to that from \citet{Guillot2010}. 
In our model, temperatures are relatively low compared with the \citet{Guillot2010} model at $P\lesssim 40$~bar, which is due to difference in opacity.
In our model, deep regions of $P \gtrsim 40$~bar are convective, while there is no convective region in the \citet{Guillot2010} model because of constant opacity. 
We have compared our $P$-$T$ profile with the \citet{Fortney2005}'s and \citet{Iro2005}'s profiles, which are shown in Fig.~6 of \citet{Guillot2010} and confirmed that our $P$-$T$ profile in the convective region is almost equal to their profiles. 
Of special interest in this study is the entropy at the radiative/convective boundary, because it governs the thermal evolution of the planet. 
In this sense, it is fair to say that our atmospheric model yields appropriate boundary conditions for the structure of the convective interior. 

Finally, we describe an analytical expression for our atmospheric model.
We basically follow the prescription developed by \cite{Heng2012}, except for the treatment of the opacity.
As \citet{Heng2012} mentioned, it would be a challenging task without assumption of constant $\kappa_{\mathrm{v}}^{\mathrm{p}}$ and $\kappa_{\mathrm{v}}^{\mathrm{r}}$ to obtain analytical solutions for $J_{\mathrm{v}}$ and $H_{\mathrm{v}}$.
Here we assume $\kappa_{\mathrm{v}}^{\mathrm{p}}$ and $\kappa_{\mathrm{v}}^{\mathrm{r}}$ are constant throughout the atmosphere.
We differentiate (\ref{RMT-01}) and (\ref{RMT-02}) by $m$ and obtain
\begin{eqnarray}
\frac{d^2 J_{\mathrm{v}}}{dm^2} &=& \frac{H_{\mathrm{v}}}{\mu^2} \frac{d\kappa_{\mathrm{v}}^{\mathrm{r}}}{dm} + \frac{\kappa_{\mathrm{v}}^{\mathrm{r}}\kappa_{\mathrm{v}}^{\mathrm{p}}}{\mu^2} J_{\mathrm{v}}, \label{RMT-01-1} \\
\frac{d^2 H_{\mathrm{v}}}{dm^2} &=&J_{\mathrm{v}}  \frac{d\kappa_{\mathrm{v}}^{\mathrm{p}}}{dm} + \frac{\kappa_{\mathrm{v}}^{\mathrm{r}}\kappa_{\mathrm{v}}^{\mathrm{p}}}{\mu^2} H_{\mathrm{v}}, \label{RMT-02-1} 
\end{eqnarray}
where $\mu^2=K_{\mathrm{v}}/J_{\mathrm{v}}$.
Assuming $J_{\mathrm{v}}=H_{\mathrm{v}}=0$ as $m\to\infty$, we obtain
\begin{eqnarray}
(J_{\mathrm{v}},~H_{\mathrm{v}}) = (J_{\mathrm{v},0},~H_{\mathrm{v},0})\exp\left(-\frac{\bar{\kappa_{\mathrm{v}}}}{\mu}m \right),
\end{eqnarray}
where $\bar{\kappa_{\mathrm{v}}} = \sqrt{ \kappa_{\mathrm{v}}^{\mathrm{p}}\kappa_{\mathrm{v}}^{\mathrm{r}} }$ and $J_{\mathrm{v},0}$ and $H_{\mathrm{v},0}$ are the values of $J_{\mathrm{v}}$ and $H_{\mathrm{v}}$ evaluated at $m=0$, respectively.
In general, the heat transportation, such as circulation, produces a specific luminosity of heat.
\citet{Heng2012} introduced the specific luminosity as $Q$, which has units of erg~s$^{-1}$~g$^{-1}$.
$Q$ can be related to the moments of the specific intensity and we obtain
\begin{equation}
\kappa_{\mathrm{th}}^{\mathrm{p}}\left(J_{\mathrm{th}} - B \right) + \kappa_{\mathrm{v}}^{\mathrm{p}} J_{\mathrm{v}} = Q. \label{RMT-05-1}
\end{equation}
We integrate Eq.~(\ref{RMT-05-1}) and obtain
\begin{equation}
H = H_{\infty} - \tilde{Q}(m,\infty) \label{RMT-05-2},
\end{equation}
where $H_{\infty}$ is the value of $H$ evaluated at $m\to\infty$ and
\begin{equation}
\tilde{Q}(m_1,m_2) = \int_{m_1}^{m_2} Q(m',\mu,\phi) dm'.
\end{equation}
To obtain $H_{\mathrm{th}}$ and $J_{\mathrm{th}}$,
we substitute Eq.~(\ref{RMT-05-1}) in Eqs.~(\ref{RMT-03}) and (\ref{RMT-04}) and integrate by $m$.
Then we obtain
\begin{eqnarray}
H_{\mathrm{th}}&= & H_{\mathrm{\infty}} -H_{\mathrm{v},0} \exp\left( -\frac{\bar{\kappa_{\mathrm{v}}} }{\mu} m \right) - \tilde{Q}(m,\infty) \\
J_{\mathrm{th}} &=& J_{\mathrm{th},0} - \frac{H_{\mathrm{v},0}}{ f_{K\mathrm{th}}}
\int_{0}^{m} \kappa_{\mathrm{th}}^{\mathrm{r}} \exp\left(-\frac{\bar{\kappa_{\mathrm{v}}}}{\mu} m' \right) dm' \nonumber \\
&&+ \frac{1}{f_{K\mathrm{th}}} \int_{0}^{m} \kappa_{\mathrm{th}}^{\mathrm{r}} \left\{ H_{\mathrm{\infty}} - \tilde{Q} (m', \infty) \right\} dm',
\end{eqnarray}
where  $f_{K\mathrm{th}}=K_{\mathrm{th}}/J_{\mathrm{th}}$, $f_{H\mathrm{th}}=H_{\mathrm{th}}/J_{\mathrm{th}}$, and
\begin{equation}
J_{\mathrm{th},0} = \frac{1}{f_{H\mathrm{th}}}\left\{ H_{\infty}-H_{\mathrm{v},0}-\tilde{Q}(0,\infty) \right\}.
\end{equation}
That is, we obtain
\begin{eqnarray}
B &=& H_{\infty}\left[ \frac{1}{f_{H\mathrm{th}}} + \frac{1}{f_{K\mathrm{th}}} \tau_{\mathrm{th}}(m) \right] \nonumber \\
&&-H_{\mathrm{v},0} \left[ \frac{1}{f_{H\mathrm{th}}} 
+\frac{\bar{\kappa_{\mathrm{v}}}}{\mu \kappa_{\mathrm{th}}^{\mathrm{p}}}
+\frac{1}{f_{K\mathrm{th}}} \tau_{\mathrm{ext}}(m) \right]
 + E(m),
\end{eqnarray}
where
\begin{eqnarray}
\tau_{\mathrm{th}}(m) &=&  \int_{0}^{m}\kappa_{\mathrm{th}}^{\mathrm{r}} dm', \\
\tau_{\mathrm{ext}}(m) &=& \int_{0}^{m}
\left( \bar{\kappa_{\mathrm{th}}}^2  - \frac{ f_{K\mathrm{th}} }{\mu^2} \bar{\kappa_{\mathrm{v}}}^2 \right)
\frac{1}{\kappa_{\mathrm{th}}^{\mathrm{p}}}\exp\left( -\frac{\bar{\kappa_{\mathrm{v}}}}{\mu}m' \right) dm' , \\
E(m) & = & -\left[ \frac{Q}{\kappa_{\mathrm{th}}^{\mathrm{p}}} + \frac{1}{f_{K\mathrm{th}}} \int_{0}^{m} \kappa_{\mathrm{th}}^{\mathrm{r}} \tilde{Q}(m',\infty)dm' + \frac{\tilde{Q}(0,\infty)}{f_{H\mathrm{th}}}  \right],
\end{eqnarray}
and $\bar{\kappa_{\mathrm{th}}} = \sqrt{\kappa_{\mathrm{th}}^{\mathrm{p}}\kappa_{\mathrm{th}}^{\mathrm{r}}}$.
In our conditions, we assume $\mu = 1/\sqrt{3}$, $f_{K\mathrm{th}}=1/3$, $f_{H\mathrm{th}}=1/2$ and $Q=0$.
Consequently, we obtain the temperature profile as
\begin{equation}
T^4 = \frac{3}{4}T_{\mathrm{int}}^4 \left[ \frac{2}{3} + \tau_{\mathrm{th}}(m)  \right]  + \frac{\sqrt{3}}{4} T_{\mathrm{irr}}^4 \left[ \frac{2}{3} + \frac{\bar{\kappa_{\mathrm{v}}}}{\sqrt{3}\kappa_{\mathrm{th}}^{\mathrm{p}}} + \tau_{\mathrm{ext}}(m) \right] \label{TP_ana}
\end{equation}
where
\begin{eqnarray}
\tau_{\mathrm{ext}}(m) &=& \int_{0}^{m} \frac{\bar{\kappa_{\mathrm{th}}}^2 - \bar{\kappa_{\mathrm{v}}}^2}{\kappa_{\mathrm{th}}^{\mathrm{p}}} \exp\left(-\sqrt{3} \bar{\kappa_{\mathrm{v}}} m' \right) dm'.
\end{eqnarray}
If we assume $\kappa_{\mathrm{th}}^{\mathrm{p}} = \kappa_{\mathrm{th}}^{\mathrm{r}} $
and $\kappa_{\mathrm{v}}^{\mathrm{p}} = \kappa_{\mathrm{v}}^{\mathrm{r}}$,
Eq.~(\ref{TP_ana}) agrees with Eq.~(27) of \citet{Heng2012}.

\begin{figure}[htbp]
\begin{center}
\resizebox{\hsize}{!}{\includegraphics{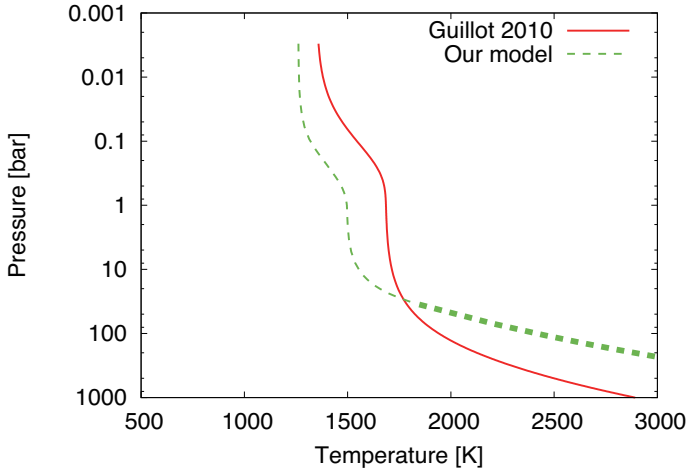}}
\caption{
Temperature-pressure profiles for a solar-composition atmosphere (see the details in text).
The solid (red) and dotted (green) lines represent the both \citet{Guillot2010}'s ($\gamma=0.4$) and our models, respectively.
The thin and thick parts of the dotted line represent the radiative and convective regions, respectively. 
We have assumed $g=980$~cm~s$^{-2}$, $T_{\mathrm{int}}=300$~K, and $T_{\mathrm{irr}}=1500$~K.
}
\label{atm_hikaku}
\end{center}
\end{figure}

\end{document}